\documentclass{emulateapj}
\usepackage{amsmath}
\usepackage{txfonts}
\usepackage{natbib}
\usepackage{graphicx}
\usepackage{color}

\newcommand{\ud}{\ensuremath{\mathrm{d}}}

\shorttitle{Neutrino flavor oscillations and neutrino-driven wind nucleosynthesis}

\shortauthors{Pllumbi, Tamborra, Wanajo, Janka,  \& H\"udepohl}
\begin{document}

\title{Impact of neutrino flavor oscillations on the neutrino-driven wind nucleosynthesis of an electron-capture supernova}

\author{Else Pllumbi\altaffilmark{1,2},
        Irene Tamborra\altaffilmark{3},
        Shinya Wanajo\altaffilmark{4},
        Hans-Thomas Janka\altaffilmark{1},
        and Lorenz H\"udepohl\altaffilmark{1}}

\altaffiltext{1}{Max-Planck-Institut f\"ur Astrophysik,
       Karl-Schwarzschild-Str.~1, D-85748 Garching, Germany;
       e-mail: epllumbi@mpa-garching.mpg.de,\\ thj@mpa-garching.mpg.de, lorenz@mpa-garching.mpg.de}

\altaffiltext{2}{Physik Department, Technische Universit\"at M\"unchen, 
James-Franck-Stra\ss e 1, 85748 Garching, Germany}

\altaffiltext{3}{GRAPPA Institute, University of Amsterdam,
Science Park 904, 1098 XH Amsterdam, The Netherlands; i.tamborra@uva.nl}

\altaffiltext{4}{iTHES Research Group, RIKEN, Wako, Saitama 351-0198, Japan;\\
e-mail: shinya.wanajo@riken.jp}

%%%%%%%%%%%%%%%%%%%%%%%%%%%%%%%%%%%%%%%%%%%%%%%%%%%%%%%%%%%%%%%%%%%%%%%%%%%%%%%%%%%%%%%%%%%%%
\begin{abstract}
Neutrino oscillations, especially to light sterile states, can
affect the nucleosynthesis yields because of their possible feedback effect on
the electron fraction ($Y_{e}$). For the first time, we perform
nucleosynthesis calculations for neutrino-driven wind trajectories 
from the neutrino-cooling phase of an 8.8 $M_{\odot}$ electron-capture 
supernova, whose hydrodynamic evolution was 
computed in spherical symmetry with sophisticated neutrino transport 
and whose $Y_{e}$ evolution was post-processed by including 
neutrino oscillations both between active and
active-sterile flavors. We also take into account the $\alpha$-effect
as well as weak magnetism and recoil corrections in the neutrino 
absorption and emission processes. We observe effects on the $Y_e$ evolution
which depend in a subtle way on the relative radial positions of the 
sterile MSW resonances, of collective flavor transformations, and on the formation of
$\alpha$ particles. 
For the adopted supernova progenitor, we find that neutrino oscillations, 
also to a sterile state with eV-mass, 
do not significantly affect the element formation
and in particular cannot make the post-explosion wind outflow neutron 
rich enough to activate a strong r-process. 
Our conclusions become even more robust
when, in order to mimic equation-of-state dependent corrections due to nucleon potential effects in the
dense-medium neutrino opacities, six cases with reduced $Y_{e}$ in the wind are considered. 
In these cases, despite the conversion of active neutrinos to sterile neutrinos, $Y_{e}$
increases or is not significantly lowered
compared to the values obtained without oscillations and active flavor 
transformations. This is a consequence of a complicated interplay between sterile-neutrino production,
neutrino-neutrino interactions, and $\alpha$-effect.

\end{abstract}


%%%%%%%%%%%%%%%%%%%%%%%%%%%%%%%%%%%%%%%%%%%%%%%%%%%%%%%%%%%%%%%%%%%%%%%%%%%%%%%%%%%%%%%%%%%%%%%%%%%%
\section{Introduction}
%%%%%%%%%%%%%%%%%%%%%%%%%%%%%%%%%%%%%%%%%%%%%%%%%%%%%%%%%%%%%%%%%%%%%%%%%%%%%%%%%%%%%%%%%%%%%%%%%%%%
Stars with mass larger than $\sim$8 $M_{\odot}$ end their life as
core-collapse supernovae~\citep[CCSNe, e.g.,][]{Woosleyetal02}.  In
particular, those with initial mass between $\sim$8 and $\sim$10 $M_{\odot}$ form
an electron-degenerate core composed of oxygen, neon, and magnesium
(O-Ne-Mg) and end their life either as O-Ne-Mg white dwarfs or as 
``electron-capture supernovae'' \citep[ECSNe,][]{Nomoto1987}, 
when electrons are captured on Ne and Mg triggering the collapse of the stellar core. 
Since ECSNe represent up to 30$\%$ of all CCSNe \citep{Ishimaru1999,
Poelarends2008, Wanajo2011aaa}, they could 
significantly contribute to the Galactic chemical enrichment with heavy elements
\citep{Ishimaru1999}.\\ 
Electron-capture supernovae were suggested as candidate sites for the
r-process (rapid neutron-capture) element production \citep{Hillebrandt1984,
Wanajo2003,Ningetal2007}. For comprehensive reviews on the r-process,
see ~\cite{Wanajo2006b}, \cite{Arnould:2007gh} and \cite{Thielemann2011}. However, recent
nucleosynthesis studies (\citealp{Hoffman2008, Wanajoetal09}), based on
self-consistent hydrodynamic simulations of the explosion \citep{Kitaura2006,
Janka2008}, do not support the production of elements with mass number heavier 
than $A\sim110$ in the early ejecta of ECSNe, but suggest interesting production of
light trans-iron elements from Zn to Zr \citep{Wanajo2011aaa}, of
$^{48}$Ca \citep{Wanajo2013} and of $^{60}$Fe \citep{Wanajoetal13b}.
Two-dimensional hydrodynamic simulations do not provide 
conditions for a strong r-process. However, a weak r-process cannot 
be excluded, if the ejecta were slightly more neutron-rich than obtained 
in the models.\\
After the launch of the SN explosion, the proto-neutron star (PNS) cools
because of the neutrino emission. Due to capture reactions and
scattering events, neutrinos deposit energy in the outer layers of the
PNS, giving birth to an outflow mainly composed of free neutrons and
protons, the so-called neutrino-driven ($\nu$-driven) wind --
see~\cite{Janka12} and \cite{Arconesetal12} for  recent reviews on the topic.  While expanding
away from the neutron star, the $\nu$-driven wind matter cools and
 nucleons recombine, producing alpha particles and some fraction of heavy nuclei. 
The $\nu$-driven wind has long been considered as a promising site of the r-process
\citep{Meyer1992, Woosley1994, Takahashi1994, Qian1996, Otsuki2000,
Wanajo2001, Thompson2001}. However, the outcome of the $\nu$-driven wind
nucleosynthesis is strongly sensitive to the electron fraction
$Y_{e}$ (number of protons per nucleon), the entropy and the expansion timescale. 
Recent long-time hydrodynamic SN simulations with elaborate neutrino
transport~\citep{Fischeretal09, Huedepohletal10} show, besides insufficient 
entropy, a trend towards 
proton-rich $\nu$-driven winds, rather than neutron-rich ones as it
would be required for an r-process to occur. Such proton-rich
conditions might be suitable for the
$\nu$\textit{p}-process making some light $p$-nuclei
\citep{Fro2005,Froelich2006a,Pruet2006, Wanajo2006a}.

More recently, however, it has been pointed out that the mean-field shift
of nucleon potential energies \citep{Reddy1998} significantly alters
the charged-current neutrino opacity in the neutrinospheric layer 
and reduces $Y_e$ from initially
proton-rich values down to possibly $\sim$0.42--0.45 for some 
temporary phase of the wind evolution \citep{Roberts2012a,
Martinez2012, Robertsetal12}. This effect was not
adequately included in previous simulations, and it  
 becomes important only when
the neutrinosphere reaches high densities (postbounce time $t_\mathrm{pb} >$~a~few~100~ms). 
At very late times, however, high
neutrinospheric densities suppress $\nu_e$ absorption on neutrons by final-state Pauli
blocking of electrons \citep{Fischer:2011cy}, $\nu_e$ escape with harder spectra, and $Y_e$ 
in the wind increases again. The
matter at early and probably late times is thus still expected to be
proton-rich.

One has to wonder whether favorable conditions for the
r-process could still occur in supernovae. Since $Y_{e}$
depends on the competition between the capture rates of $\nu_e$ and
$\bar{\nu}_e$ on free nucleons and their inverse
reactions~\citep{FullerMeyer95}, a modification of the predicted 
neutrino energy spectra, for example due to nucleon-potential 
effects, could affect $Y_e$ in the neutrino-driven outflows. 
Moreover, neutrino flavor oscillations could  modify the wind-$Y_{e}$,
if they significantly alter the $\nu_e$ and $\bar{\nu}_e$ fluxes 
before $Y_e$ reaches its asymptotic value.
Therefore, the inclusion of flavor oscillations may be
crucial for determining the nuclear production in the $\nu$-driven wind
matter and to clarify whether ECSNe could still be considered as
candidate sites for the r-process.

The nucleosynthesis yields (and the r-process) in supernovae
might be affected by the existence of light sterile neutrinos, hypothetical
gauge-singlet fermions that could mix with one or more of the active
states and thus show up in active-sterile flavor
oscillations (see~\citealp{Abazajianetal12,Palazzo:2013} for recent reviews on the 
topic). In particular, eV-mass sterile
neutrinos with large mixing imply that the $\nu_e$ flux would undergo
Mikheyev-Smirnov-Wolfenstein (MSW)
conversions~\citep{wolf,Wolfenstein:1977ue} to $\nu_s$ closer to the SN
core than any other oscillation effect. We assume that the sterile state is heavier
 than the active ones because of cosmological neutrino mass limits~\citep{Abazajianetal12}. 
 The idea that removing the $\nu_e$ flux by active-sterile oscillations
could favor a neutron-rich outflow environment was proposed some time
ago \citep{Beun:2006ka,Keranen:2007ga,Fetter:2002xx,Fetter:2000kf,
McLaughlinetal99,Hidaka:2007se,Nunokawa:1997ct}. However, the considered mass
differences were larger and the possible impact of $\nu$-$\nu$ interactions in the
active sector~\citep{Duan:2010bg} was not taken into
account.  

Recently, low-mass sterile neutrinos have
been invoked to explain the excess $\bar{\nu}_e$ events in the LSND
experiment \citep{Aguilar:2001ty,Strumia:2002fw,GonzalezGarcia:2007ib}
as well as the MiniBooNE
excess~\citep{AguilarArevalo:2009xn,AguilarArevalo:2008rc,Karagiorgi:2009nb,AguilarArevalo:2012va}.
Moreover an indication for the possible existence of eV-mass sterile
neutrinos comes from a new analysis of reactor $\bar\nu_e$ spectra and
short-baseline experiments~(\citealp{Kopp:2011qd,Giunti:2011cp,Giunti:2011hn,Giunti:2012tn,Donini:2012tt,Giunti2013PhRvD,Kopp:2013vaa}). 
The cosmic microwave background
anisotropies~\citep{Reid:2009nq,Hamann:2010bk,Hou:2011ec,Hinshaw:2012fq,Ade2013lta,Archidiacono:2013xxa}
point towards a cosmic excess radiation compatible with one family of
fully thermalized sub-eV sterile neutrinos or one or even two partially thermalized
sterile neutrino families with sub-eV/eV mass~\citep{Archidiacono:2013xxa,Giusarma2014,Archidiacono:2014apa}.

Such intense activity triggered new interest in the role of 
neutrino oscillations with and without sterile neutrinos, and including $\nu$-$\nu$ 
interactions, on nucleosynthesis processes like the r-process and the
$\nu$\textit{p}-process in SN outflows~\citep{Tamborraetal12,
Duan:2010af,MartinezPinedoetal11,Wu:2014kaa}.

The role of active-sterile neutrino mixing 
for the $\nu$-driven explosion mechanism and the nucleosynthesis in the 
early ($t \leq 100$~ms postbounce) ejecta of ECSNe was discussed by \cite{Wuetal2013}.
The authors found that active-sterile conversions can not only suppress
neutrino heating considerably but also potentially  
enhance the neutron-richness of the ejecta allowing 
for the production of the elements from
Sr, Y and Zr up to Cd. The conclusiveness of these results is unclear,
however, because, besides approximate modeling of neutrino oscillations,  
only spherically symmetric models were considered, although
multi-dimensional effects had been shown to be important during 
the onset of the explosion (cf. \citealp{Wanajo2011aaa}). In contrast to spherical models, 
multi-dimensional ones provide sufficient neutron excess to 
yield interesting amounts of elements between the Fe-group 
and $\mathrm{N} \,=\, 50$ nuclei even without involving sterile neutrino effects \citep{Wanajo2011aaa}.

In this work, we explore the impact of neutrino flavor oscillations
(with and without the inclusion of an extra eV-mass sterile neutrino) on
the $Y_{e}$ evolution of the $\nu$-driven wind and on the corresponding
nucleosynthesis yields
of an ECSN, whose evolution can be well described in spherical symmetry 
and has been followed beyond the
explosion continuously into the subsequent proto-neutron star cooling phase
\citep{Huedepohletal10}. The simulation of
\cite{Huedepohletal10} did not include the aforementioned nucleon mean-field
effects in the charged-current neutrino-nucleon reactions and 
resulted in the ejection of proton-rich matter throughout
the wind phase. We still use this model to examine neutrino
oscillation effects in the neutrino-driven wind, because the 
wind dynamics and thermodynamics conditions are only marginally 
changed despite the impact of the nucleon potentials on the 
electron fraction (e.g.,~\citealp{Martinez2012}).

Our paper is structured in the following way.
In Sect.~2, we describe the $\nu$-driven wind
trajectories adopted for the nucleosynthesis calculations, as well as
our reaction network. In Sect.~3, the electron fraction evolution and the
nucleosynthesis results are presented when no neutrino oscillations occur as 
fiducial case. After introducing the neutrino mass-mixing
parameters in Sect.~4, we briefly discuss the oscillation physics involved in the
nucleosynthesis calculations in Sect.~5. Our results for $Y_{e}$ and how it
is affected by neutrino oscillations (with and without sterile
neutrinos) including the corresponding nucleosynthesis are presented in Sect.~6.  
In Sect.~7, we introduce six toy model cases for the ${\nu}_e$ and $\bar{\nu}_e$ energy spectra
in order to explore the possible consequences of nuclear mean-field effects
in the neutrino opacities. In Sect.~8, we discuss our results and compare 
with other works. We present our conclusions and perspectives in Sect.~9. In an Appendix,
we give more details about the feedback of neutrino self-interactions on $Y_{e}$.

%%%%%%%%%%%%%%%%%%%%%%%%%%%%%%%%%%%%%%%%%%%%%%%%%%%%%%%%%%%%%%%%%%%%%%%%%%%%%%%%%%%%%%%%%%%%%
\section{Neutrino-driven wind and reaction network}
%%%%%%%%%%%%%%%%%%%%%%%%%%%%%%%%%%%%%%%%%%%%%%%%%%%%%%%%%%%%%%%%%%%%%%%%%%%%%%%%%%%%%%%%%%%%%
We use one-dimensional (1D) long-time
simulations of a representative $8.8\,M_{\odot}$
progenitor~\citep{Huedepohletal10}, performed with the equation of state
of~\cite{Shenetal98}. For the present study we adopt the Model Sf~21
(see~\citealp{Huedepohletal10} for further details\footnote{Model Sf~21 is analog to
model Sf of \cite{Huedepohletal10} but was computed with 21 energy bins for the neutrino
transport instead of the usual 17 energy groups.}). In the chosen model,
the accretion phase ends already at a postbounce time of $t_\mathrm{pb}\sim0.2$~s when
neutrino heating drives the expansion of the postshock layers and powers 
the explosion. The subsequent deleptonization and cooling of the PNS were 
followed for $\sim10$~s.

\begin{figure*}
\resizebox{0.9\textwidth}{!}{
\includegraphics*{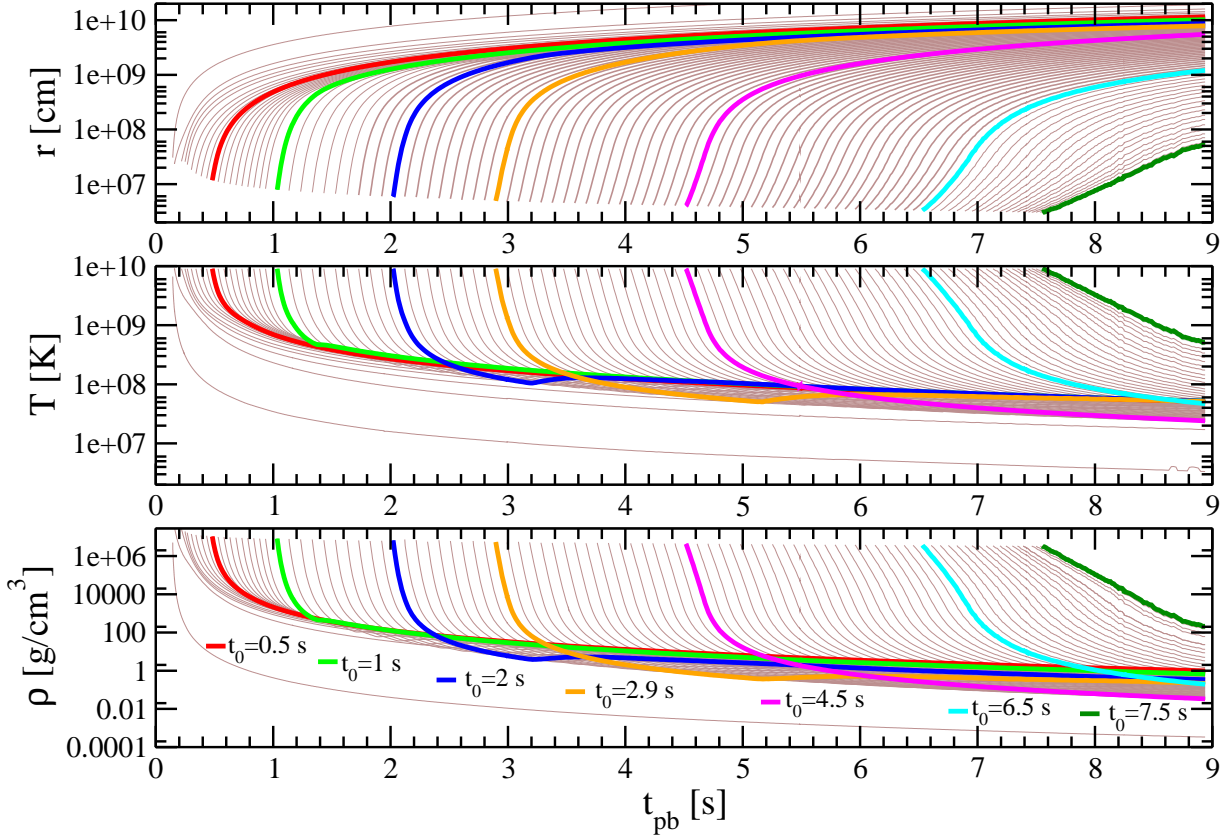}}
\caption{Mass-shell trajectories of the neutrino-driven wind as functions of postbounce time 
($t_{{\mathrm{pb}}}$): Radial distance from the PNS center (top), temperature (middle), 
and density (bottom) along the ejecta trajectories.
The colored curves correspond to the selected 7 trajectories representative of the 
evolution of the $\nu$-driven wind at initial times $t_0 = 0.5, 1, 2, 2.9, 4.5, 6.5, 7.5$~s.
The kinks visible in the temperature and density evolutions of the trajectories at $t_0 = 2$~s 
and $2.9$~s indicate the existence of a weak reverse shock.}
\label{fig:Rad_Temp_Rho}
\end{figure*}

In order to perform the network calculations for the nucleosynthesis in the 
neutrino-driven wind, we use 98 ejecta trajectories.
Figure~\ref{fig:Rad_Temp_Rho} shows the time evolution of the distance $r$ from the center of
the PNS (top panel), temperature $T$ (middle panel), and
 matter density $\rho$ (bottom panel) for these mass-shell trajectories as functions of $t_{\mathrm{pb}}$. 
 %--------------------------------------------
%--------------------------------------------
The outflow evolution of 7 of the 98 trajectories,
corresponding to initial times $t_0 = 0.5, 1, 2, 2.9, 4.5, 6.5, 7.5$~s ($t_0 $ being
measured when the temperature $T_0=9$ GK), 
is highlighted with different colors. We adopt these seven trajectories as representative of 
the cooling evolution of the PNS
to discuss the impact of neutrino oscillations (with and without an
additional light sterile neutrino) on the nucleosynthesis in the $\nu$-driven wind. 
The total ejecta mass of the 98 mass-shell trajectories is $M_{98}=1.1
\times 10^{-2} M_{\odot}$. 

In the network, 6300 species are included between the proton-drip line 
and neutron-drip line, up to the $Z=110$ isotopes~\citep[see][for more details]{Wanajoetal09}. 
All the important reactions such as $\nu_e(n,p)e^{-}$, $\bar\nu_e(p,n)e^{+}$, $(n,\gamma)$, $(p,\gamma)$,
$(\alpha,\gamma)$, $(p,n)$, $(\alpha,n)$, $(\alpha,p)$, and their
inverse ones are taken into account. 
The $\nu_{e}$ and $\bar\nu_{e}$ capture rates on free neutrons and
protons are calculated as in~\cite{HorowitzLi99} and thus include recoil 
and weak magnetism corrections.
The neutrino-induced reactions on heavy nuclei are
not included since they have negligible effects \citep{Meyer1998}. The nucleosynthesis
calculations start when the mass-shell temperature decreases to
$9$ GK, with an initial composition of free neutrons and protons with 
number fractions of $1-Y_e$ and $Y_e$, respectively.

%%%%%%%%%%%%%%%%%%%%%%%%%%%%%%%%%%%%%%%%%%%%%%%%%%%%%%%%%%%%%%%%%%%%%%%%%%%%%%%%%%%%%%%%%%%%%%
\section{Electron fraction evolution}    \label{sec:yeevo}
%%%%%%%%%%%%%%%%%%%%%%%%%%%%%%%%%%%%%%%%%%%%%%%%%%%%%%%%%%%%%%%%%%%%%%%%%%%%%%%%%%%%%%%%%%%%%%
The matter in a fluid element moving away from the PNS will experience
three stages of nuclear evolution. Near the surface of the PNS, the temperature is so
high that the matter is in nuclear statistical equilibrium (NSE) and nearly all of
the baryons are in the form of free nucleons. As the
material flows away from the PNS, it cools. When the temperature is $T <
1$~MeV, $\alpha$ particles begin to assemble to form heavier nuclei 
 by $\alpha$$\alpha$$n$, 3$\alpha$ reactions,
and subsequent captures of $\alpha$ particles and free nucleons. 

Together with the entropy and the expansion time, a basic
quantity defining the conditions for element formation 
(and eventually the r-process) is the excess of initially free $n$ or $p$ 
expressed by the electron fraction $Y_{e}$.  
It is locally defined as the ratio of the net electron (electrons minus
positrons) number density, $N_e$, to the sum of proton number density $N_p$
and neutron number density $N_n$:

%....................................................
\begin{equation}
\label{yedef}
\hspace{-0.1mm}
Y_e(r)=\frac{N_e(r)}{N_p(r)+N_n(r)}=X_{p}(r)+\frac{X_{\alpha}(r)}{2}+\sum_{Z_A>2}{\frac{Z_A(r)}{A(r)}X_A(r)}\ ,
\end{equation}
%.................................................... 
where $X_{p}$, $X_{\alpha}$, and $X_A$ are the mass fractions of free protons
($p$), $\alpha$ particles, and heavy elements ($Z_A > 2$) as functions of
the radius. The charge and the mass numbers of the heavy nuclear 
species are $Z_A$ and $A$, respectively. In all neutral media, $Y_e=Y_p$ and $Y_n=1-Y_e$,
with $Y_j$ being the number density of free or bound particle species $j$ relative to
baryons. The lower $Y_{e}$ is, the more the environment is neutron
rich, and thus the more favorable it is for the r-process to occur 
(e.g.,~\citealp{Hoffmanetal97}).
On the other hand, $Y_{e}>0.5$ implies that $p$-rich nuclei could be formed through the ${\nu}p-$process
\citep{Fro2005, Pruet2006, Wanajo2006a}.

Having in mind the overall evolution of abundances with radius and time
and assuming that the reactions of neutrinos on nuclei are negligible,
the $n/p$ ratio in the wind ejecta is set by $\beta$-interactions 
of electron neutrinos ($\nu_{e}$) and
electron antineutrinos ($\bar\nu_{e}$) with free $n$ and $p$ and
their inverse reactions:
%....................................................
\begin{eqnarray}
\label{capture1}
\nu_e + n &\rightleftharpoons& p + e^-\ ,\\
\label{capture2}
\bar{\nu}_e + p  &\rightleftharpoons& n + e^+.
\end{eqnarray}
%....................................................
Therefore the $Y_{e}$ evolution  depends on the energy distributions  
of $\nu_{e}$ and $\bar\nu_{e}$. Modifications of the neutrino 
emission properties, such as the energy spectra, due to flavor oscillations could significantly 
change the $n/p$ ratio and thus $Y_{e}$ in the wind.

Because of slow time variations of the outflow
conditions during the PNS cooling phase, a near steady-state situation
applies \citep{Qian1996} 
and the rate-of-change of $Y_e$ within an outflowing mass element
can be written as in~\cite{McLaughlinetal96}:
%....................................................
\begin{equation}
\label{Yeeq}
\frac{dY_e}{dt} = v(r) \frac{dY_e}{dr} \simeq (\lambda_{\nu_e} + \lambda_{e^+}) Y_n^{\rm f} - (\lambda_{\bar{\nu}_e} + 
\lambda_{e^-}) Y_p^{\rm f}\ ,
\end{equation}
%....................................................
with $v(r)$ being the velocity of the outflowing mass element,  $\lambda_{i}$ 
the reaction rates, and $Y_{n, p}^{\rm f}$  the abundances of free nucleons.

In the free streaming limit with neutrinos propagating radially,
the forward reaction rates of Eqs.~(\ref{capture1},\ref{capture2}) can be 
written in terms of the electron (anti)neutrino emission properties:
%....................................................
\begin{eqnarray}
\label{lambdanue}
\lambda_{\nu_e} &\simeq& \frac{L_{\nu_e}}{4 \pi r^2 \langle E_{\nu_e}\rangle}\, \langle \sigma_{\nu_e}\rangle \ ,
\\
\label{lambdaantinue} \lambda_{\bar{\nu}_e} &\simeq&
\frac{L_{\bar{\nu}_e}}{4 \pi r^2 \langle E_{\bar{\nu}_e} \rangle}\,
\langle \sigma_{\bar{\nu}_e}\rangle \ ,
\end{eqnarray}
%....................................................
where $L_{\nu_{e}}$ and $L_{\bar{\nu}_e}$ are the luminosities of
$\nu_e$ and $\bar{\nu}_e$ respectively, $\langle E_{\nu_e} \rangle$
and $\langle E_{\bar{\nu}_e} \rangle$ the mean spectral energies\footnote{$\langle E^n_{\nu} \rangle$ $\equiv \int E^n_{\nu}f(E_{\nu}) \,\ud E$, 
where $f(E_{\nu})$ is the normalized (anti)neutrino energy spectrum. The energy spectrum which we use will be described in Sect.~4.}.
The $\nu_{e}$ and $\bar{\nu}_e$ capture cross sections of the forward reactions
~(\ref{capture1},\ref{capture2}), averaged over the corresponding $\nu_{e}$ and $\bar{\nu}_e$ energy spectra, 
are $\langle{\sigma_{\nu_e}}\rangle$ and $\langle \sigma_{\bar{\nu}_e} \rangle$, respectively.
Including the weak magnetism and recoil corrections, 
the average neutrino capture cross sections
are~\citep{HorowitzLi99}:
%....................................................
\begin{eqnarray}
\label{sigmae}
\hspace*{-0.5cm}\langle\sigma_{\nu_e}\rangle&\simeq&k {\left\langle{E}_{\nu_e}\right\rangle}\varepsilon_{\nu_e}\left[1+2 
\dfrac{\Delta}{\varepsilon_{\nu_e}} +
a_{\nu_e}\left( {\dfrac{\Delta}{\varepsilon_{\nu_e}}}\right)^{2}\right] W_{\nu_e},
\\  
\hspace*{-0.5cm}\langle\sigma_{\bar\nu_e}\rangle&\simeq&k {\left\langle{E}_{\bar\nu_e}\right\rangle}\varepsilon_{\bar\nu_e}\left[1-2 
\dfrac{\Delta}{\varepsilon_{\bar\nu_e}}+a_{\bar{\nu}_e}\left( 
{\dfrac{\Delta}{\varepsilon_{\bar\nu_e}}}\right)^{2}\right] W_{\bar{\nu}_e},
\label{sigmaanue}
\end{eqnarray}
%....................................................
with $k\simeq9.3 \times 10^{-44}$ cm$^{2}$/MeV$^{2}$, 
$\varepsilon_{\nu} = \langle{E}^{2}_{\nu}\rangle/\langle{E}_{\nu}\rangle$ ($\nu = \nu_e, \bar{\nu_e}$), 
 $a_{\nu}=\langle{E^2_{\nu}}\rangle/{\langle{E_{\nu}}\rangle}^{2}$,
$M$  the nucleon mass in MeV, and $\Delta=1.293$ MeV the neutron-proton mass difference.
The weak magnetism and recoil correction factors are given by 
$W_{\nu_e}=\left[1+1.02b_{\nu_e}\varepsilon_{\nu_e}/M 
\right]$ and $W_{\bar{\nu}_e}=\left[1-7.22b_{\bar{\nu}_e} \varepsilon_{\bar\nu_e}/M \right]$, 
where $b_{\nu}=\langle{E^3_{\nu}}\rangle\langle{E_{\nu}}\rangle/{\langle{E^2_{\nu}}\rangle}^{2}$ represents 
the spectral shape factor for $\nu_{e}$ or $\bar{\nu}_e$. 
We point out that in Eq.~(\ref{sigmaanue}) the spectral integration was 
approximated by integrating over the interval $[0,\infty)$ instead of $[\Delta,\infty)$. 
Since the rates $\lambda_{\nu_e}$ and $\lambda_{\bar{\nu}_e}$ are functions of the neutrino fluxes,
they can be affected by neutrino flavor conversions.

The inverse reaction rates of ~(\ref{capture1},\ref{capture2}), $\lambda_{e^-}$ and $\lambda_{e^+}$, are defined
in analogy to the forward reaction rates: 
\begin{equation}
{\lambda}_{e^-}= c \cdot \widetilde{n}_{e^-}\cdot \langle \sigma_{e^-} \rangle, 
\label{cap_ele}
\end{equation}
\begin{equation}
\mathcal{\lambda}_{e^+}=c \cdot {n}_{e^+}\cdot \langle \sigma_{e^+} \rangle, 
\label{cap_pos}
\end{equation}
being $c$  the speed of light.
In Eq.~(\ref{cap_ele}), $\widetilde{n}_{e^-}$ is slightly modified compared
to the electron number density 
\begin{equation}
\widetilde{n}_{e^-}=\frac{8 \pi}{(2\pi \hbar c)^3}  \cdot \int\limits_0^\infty
\frac{\epsilon^2}{1+\exp
[\frac{\epsilon-\widetilde{\mu}_e}{k_{B}T}]} \,\ud \epsilon ,
\end{equation}
with $\widetilde{\mu}_e=\mu_e-\Delta$ and $\mu_e$ the electron chemical potential.
The average cross section $\langle \sigma_{e^-} \rangle$ of the inverse reaction (2) 
is
\begin{equation}
 \langle \sigma_{e^-} \rangle \simeq \frac{1}{2}k\langle \widetilde{E}_{e^-} \rangle \varepsilon_{e^-}\left[1+2 
\dfrac{\Delta}{\varepsilon_{e^-}} +
a_{e^-}\left( {\dfrac{\Delta}{\varepsilon_{e^-}}}\right)^{2}\right] W_{\nu_e},
\label{sigma_ele}
\end{equation}
where $\varepsilon_{e^-}=\langle \widetilde{E}^{2}_{e^-}\rangle/\langle \widetilde{E}_{e^-}\rangle$
and $a_{e^-}=\langle \widetilde{E}^{2}_{e^-}\rangle/{\langle \widetilde{E}_{e^-}\rangle}^2$.
In analogy to $\langle E^{n}_{\nu}\rangle$, $\langle \widetilde{E}^{n}_{e^-}\rangle$ is defined by using 
${\widetilde{f}_{e^-}}(E)=\frac{\widetilde{\xi} \cdot E^2}{1+\exp{[(E-\widetilde{\mu}_e)/k_{B}T]}}$ for the 
electron distribution function with $\widetilde{\xi}$ the normalization factor such that 
$\int \widetilde{f_e}(E)\,\ud E =1$. 
In Eq.~(\ref{cap_pos}), the positron number density is 
\begin{equation}
{n}_{e^+}=\frac{8 \pi}{(2\pi \hbar c)^3}  \cdot \int\limits_0^\infty
\frac{\epsilon^2}{1+\exp
[\frac{\epsilon+\mu_e}{k_{B}T}]} \,\ud \epsilon ,
\end{equation}
and the positron average capture cross section is defined in the following way:
\begin{equation}
 \langle \sigma_{e^+} \rangle \simeq \frac{1}{2}k\langle {E}_{e^+} \rangle \varepsilon_{e^+}\left[1+2 
\dfrac{\Delta}{\varepsilon_{e^+}} +
a_{e^+}\left( {\dfrac{\Delta}{\varepsilon_{e^+}}}\right)^{2}\right] W_{\bar{\nu}_e},
\label{sigma_pos}
\end{equation}
being $\varepsilon_{e^+}=\langle {E}^{2}_{e^+}\rangle/\langle {E}_{e^+}\rangle$
and $a_{e^+}=\langle {E}^{2}_{e^+}\rangle/{\langle {E}_{e^+}\rangle}^2$.
The energy moments are calculated using the positron distribution function
$f_{e^{+}}(E)=\frac{{\xi_{e+}} \cdot E^2}{1+\exp{[(E-{\mu}_{e^+})/k_{B}T]}}$, where
${\xi}_{e^+}$ is the normalization factor such that $\int {f_{e^{+}}}(E)\,\ud E =1$. \\
We approximate the weak magnetism and recoil corrections in Eqs.~(\ref{sigma_ele},\ref{sigma_pos})
by using $W_{\nu_e}$ and $W_{\bar{\nu}_e}$ of Eqs.~(\ref{sigmae},\ref{sigmaanue}) with
the energy moments of the neutrinos produced by the $e^{+}$ and $e^{-}$ capture reactions, 
fulfilling the detailed balance theorem.
In Eqs.~(\ref{sigma_ele},\ref{sigma_pos}) and Eqs.~(\ref{sigmae},\ref{sigmaanue}) 
we have neglected the mass of the electron, $m_e$, since it does not make any difference 
in our calculations ($m_e \ll E \pm \Delta$). 
The rates ${\lambda}_{e^-}$ and ${\lambda}_{e^+}$ are given in \cite{Bruenn85}, neglecting
weak magnetism and recoil corrections (i.e., for $W_{\nu_e}=W_{\bar{\nu}_e}=1$), 
but including $m_e$-dependent terms.

The nucleons involved in the $\beta$-reactions of Eq.~(\ref{Yeeq}) are free. 
Accounting for the nucleons bound in $\alpha$ particles, the number fractions of free protons
and neutrons can be written as functions of $Y_{e}$:
%....................................................
\begin{eqnarray}
\label{Ypf}
Y_p^{\rm f} &=& Y_e - \frac{X_\alpha}{2} -\sum_{Z_A>2} \frac{Z_A}{A} X_A\ ,\\
Y_n^{\rm f} &=& 1 - Y_e - \frac{X_\alpha}{2} -\sum_{Z_A>2} \frac{N_A}{A} X_A\ ,
\label{Ynf}
\end{eqnarray}
%....................................................
where $X_\alpha$ ($X_A$) is the mass fraction of $\alpha$ particles
(heavy nuclei). In Table~1, we list the $Y_e$ values at
the neutrinosphere\footnote{The neutrinosphere is defined as the 
region at which the neutrinos or antineutrinos escape from the 
proto-neutron star surface. We notice that, in general, 
the neutrinosphere $R_\nu$ is different for different 
(anti)neutrino flavors. We assume $R_\nu$ to be roughly the same for 
all flavors.} radius $R_\nu$ 
for the selected seven postbounce times $t_0$, as obtained from the numerical 
simulation of Model Sf 21 of \cite{Huedepohletal10}.

%-------------------------------------------------------------------------------
\begin{deluxetable*}{cccccccccccccc}
\tablecolumns{13}
\tabletypesize{\scriptsize}
\tablecaption{
Neutrinospheric parameters and electron fractions $Y_e$ as functions of
postbounce time $t_0$.
\label{table:models_res_1}
}
\tablewidth{0pt}
\tablehead{
  \colhead{$t_{0}$\tablenotemark{a}}                                     &
  \colhead{$R_\nu$\tablenotemark{b}}                                     &
  \colhead{$Y_e$\tablenotemark{c}}                                       &
  \colhead{${Y}_{e,\mathrm{a}}$\tablenotemark{d}}                                 &
  \colhead{$\Delta{\overline M}_j$ \tablenotemark{e}}                                 &
  \colhead{$L_{\nu_e}$\tablenotemark{f}}                                 &
  \colhead{$L_{\bar{\nu}_e}$\tablenotemark{g}}                           &
  \colhead{$L_{\nu_x}$\tablenotemark{h}}                                 &
  \colhead{$\langle E_{\nu_e} \rangle$\tablenotemark{i}}                 &
  \colhead{$\langle E_{\bar{\nu}_e} \rangle$\tablenotemark{j}}           &
  \colhead{$\langle E_{\nu_x} \rangle$\tablenotemark{k}}                 &
  \colhead{$\alpha_{{\nu}_e}$\tablenotemark{l}}                                  &
  \colhead{$\alpha_{\bar{\nu}_e}$\tablenotemark{m}}                      &
  \colhead{$\alpha_{\nu_x}$\tablenotemark{n}}                                  \\
  \colhead{[s]}                                                          &
  \colhead{[$10^5$ cm]}                                                           &
  \colhead{}                                                          &
  \colhead{}                                                          &
  \colhead{[10$^{-3}$ M$_\odot$]}                                                          &
  \colhead{[10$^{51}$erg/s]}                                            &
  \colhead{[10$^{51}$erg/s]}                                            &
  \colhead{[10$^{51}$erg/s]}                                            &
  \colhead{[MeV]}                                                        &
  \colhead{[MeV]}                                                        &
  \colhead{[MeV]}                                                             
   }
\startdata
0.5 & 25.0 & 0.0547 &  0.554 & 9.640 & 9.5  & 10.10  & 10.80  & 16.8  & 18.1  & 18.3 & 2.9 & 3.0 & 2.8 \\ 
1.0 & 20.5 & 0.0522 &  0.546 & 0.770 & 7.3  &  8.30  &  7.90  & 15.9  & 17.4  & 17.3 & 3.0 & 2.9 & 2.6 \\
2.0 & 17.5 & 0.0445 &  0.564 & 0.380 & 4.7  &  4.90  &  5.30  & 15.3  & 16.5  & 16.1 & 3.2 & 2.7 & 2.3 \\
2.9 & 16.0 & 0.0323 &  0.566 & 0.110 & 3.3  &  3.40  &  3.70  & 15.8  & 16.3  & 15.7 & 3.1 & 2.3 & 2.5\\
4.5 & 15.2 & 0.0268 &  0.574 & 0.060 & 1.9  &  1.90  &  2.00  & 13.8  & 13.4  & 12.9 & 3.0 & 2.3 & 2.1\\
6.5 & 14.5 & 0.0233 &  0.555 & 0.020 & 1.0  &  0.99 &   1.04  & 12.4  & 11.9  & 11.8 & 2.6 & 2.3 & 2.4\\
7.5 & 14.5 & 0.0223 &  0.549 & 0.002 & 0.6  &  0.60  &  0.60  &  9.9  &  9.6  & 9.5  & 2.4 & 2.3 & 2.5
\enddata
\tablenotetext{a}{Postbounce time.}
\tablenotetext{b}{Neutrinosphere radius.}
\tablenotetext{c}{Electron fraction at $R_\nu$.}
\tablenotetext{d}{Asymptotic electron fraction (at $r=3\times 10^7$ cm).}
\tablenotetext{e}{$\Delta{\overline M}$: ejecta mass of the 7 representative wind trajectories.}
\tablenotetext{f, g, h}{Luminosities of $\nu_e, \bar{\nu}_e$ and $\nu_x$, respectively.}
\tablenotetext{i, j, k}{Mean energies of $\nu_e, \bar{\nu}_e$ and $\nu_x$, respectively.}
\tablenotetext{l, m, n}{Spectral fitting parameters of $\nu_e, \bar{\nu}_e$ and $\nu_x$, 
respectively (see Sect.~4).}
\end{deluxetable*}
%-------------------------------------------------------------------------------
Since we aim to discuss the role of neutrino oscillations and of the so-called ``$\alpha$-effect'' on
the electron fraction and on the nucleosynthesis in the $\nu$-driven wind, we distinguish two cases with different 
$X_\alpha$ in what follows:

\begin{itemize}
 \item[(i)]We compute $X_\alpha$ using the full network (labelled ``incl. $\alpha$-effect'');
 \item[(ii)] We keep $X_{\alpha}$ constant at its value at $T = 9$~GK 
             as given by Model~Sf~21.
\end{itemize} 
The recombination of free nucleons to $\alpha$ particles affects $Y^\mathrm{f}_{p}$ and $Y^\mathrm{f}_{n}$
according to Eqs.~(\ref{Ypf}) and (\ref{Ynf}) and via Eq.~(\ref{Yeeq}) influences the evolution of $Y_{e}$.
Since the formation of $\alpha$ particles  binds equal numbers of neutrons and protons,
the remaining free nucleons will be dominated by the more abundant nucleonic species, 
either $n$ or $p$. The corresponding capture reactions of $\nu_e$ (and $e^+$)
on neutrons in the case of neutron excess or of $\bar{\nu}_e$ (and $e^-$) on protons for 
proton-rich conditions will drive $Y_{e}$ closer to 0.5, which is the so-called $\alpha$-effect
first pointed out by \cite{McLaughlinetal96} and \cite{Meyer1998}.
Since a proper inclusion of the $\alpha$-effect always requires detailed 
network calculations as in our case (i), we  consider case (ii) %, i.e.~$X_{\alpha}$  constant  for $T< 9$~G,
for isolating the effect of the formation of $\alpha$ particles on $Y_{e}$, as we will  
elucidate in Sect.~\ref{sec:Ye_Xa}.

%%%%%%%%%%%%%%%%%%%%%%%%%%%%%%%%%%%%%%%%%%%%%%%%%%%%%%%%%%%%%%%%%%%%%%%%%%%%%%%%%%%%%%%%%%%%%%%%%%%%%%%%
\subsection{Nucleosynthesis yields without neutrino oscillations} \label{subsec:no_oscill_nucl}
%%%%%%%%%%%%%%%%%%%%%%%%%%%%%%%%%%%%%%%%%%%%%%%%%%%%%%%%%%%%%%%%%%%%%%%%%%%%%%%%%%%%%%%%%%%%%%%%%%%%%%%%
\cite{Wanajo2011aaa} studied in detail the nucleosynthesis yields 
during the first 250-300 ms of the explosion of an $8.8=M_{\odot}$ ECSN in 1D and 2D. 
In this section, we discuss as our fiducial case the results of nucleosynthesis in the subsequent
$\nu$-driven wind ejecta of the same model without taking into 
account neutrino oscillations (but including the $\alpha$-effect).
Note that nucleosynthesis computations were done in previous papers  adopting
semi-analytically \citep{Wanajo2001, Wanajo2006a} or hydrodynamically
\citep{Froelich2006a, Takahashi1994, Pruet2006,Arcones:2010yt} computed 
neutrino-driven winds. With the exception of investigations by
\cite{Meyer1992} and \cite{Woosley1994}, who used the now outdated model of J. Wilson, however, 
the other existing calculations were based on a
number of simplifications or considered only constrained periods of evolution (like \citealp{Pruet2006}). 
In this sense, our study is the first one in
which the wind nucleosynthesis is explored in a self-consistently
exploded progenitor, whose evolution was continuously followed from collapse to beyond
the explosion through the complete subsequent proto-neutron star cooling
phase. Nevertheless, the results should not be taken as firm
nucleosynthetic prediction to be used for galactic chemical evolution
studies because of the absence of dense-medium nucleon potential effects 
in the charged-current neutrino reactions of the hydrodynamic
simulation. The inclusion of these nucleon-potential effects will 
cause nuclear equation-of-state dependent modifications of the neutrino emission 
and therefore of the $Y_{e}$ evolution in the $\nu$-driven wind (e.g 
\citealp{Martinez2012, Roberts2012a, Robertsetal12}), whose
investigation is beyond the present work.

Taking into account 98 trajectories, $X_{A}$ is given by
%--------------------------------------------------------------------------------
\begin{equation}
X_{A} = \frac{1}{M_{\mathrm{tot}}} \sum^{98}_{i=1}X_{i,A}\, \Delta M_{i}\ , 
\label{eq:X98_A}
\end{equation}
%--------------------------------------------------------------------------------
where $X_{i,A}$ and $\Delta M_{i}$ are the mass fractions and the ejecta-shell 
masses respectively, while $M_{\mathrm{tot}}$ is the total mass of
the ejecta, which we consider to be the sum of the ejected mass from 
the core plus the outer H/He-envelope (assumed to contain no heavy elements):
%--------------------------------------------------------------------------------
$$
M_{\mathrm{tot}} = (8.8\ M_{\odot}-1.366\ M_{\odot}) + 0.0114\ M_{\odot} \simeq 7.44\ M_{\odot}\ .
$$
%--------------------------------------------------------------------------------
Here 1.366 $M_{\odot}$ defines the initial mass cut between neutron star and ejecta and 
$M_{98}=0.0114\ M_{\odot}$.
In order to discuss the impact of neutrino oscillations\footnote{We assume that the 
$\nu_e$ and $\bar\nu_e$ luminosities and energy spectra do not change for $r\geq R_\nu$.
This means that we do not only ignore small evolutionary changes due to 
remaining neutrino interactions in the external medium but we also disregard 
general relativistic redshift corrections, which depend on $r$, and 
which are included in the hydrodynamic simulations.} in the
following sections, we replace the full set of 98 trajectories by 7
``representative'' $\nu$-driven wind trajectories
(Fig.~\ref{fig:Rad_Temp_Rho}).

For the 7 representative wind trajectories, 
we define combined mass elements, $\Delta\overline{M}_{j}$ 
($j=1,...,7$), in such a way that 
$\Delta\overline{M}_{j}={\sum}_{i=i_{j-1}+1}^{i_{j}}\Delta M_i$,
where the summation includes all mass shells ejected between the 
representative shell $i_{j-1}$ and the representative shell $i_{j}$ (see Table~1).
The first representative shell, for example, includes all the 10 
trajectories of the full set which are ejected before $t_{0}=0.5$~s.
Thus, for the 7 representative trajectories, we define
%--------------------------------------------------------------------------------
\begin{equation}
\overline{X}_{A} = \frac{1}{M_{\mathrm{tot}}} \sum^{7}_{j=1} X_{j,A}\, \Delta\overline{M}_{j}\ ,
\label{eq:X_A}
\end{equation}
%--------------------------------------------------------------------------------
with $X_{j}$ being the mass fractions for the $j$-th trajectory. 
%--------------------------------------------
\begin{figure*}
\vspace{0.8cm}
%\epsscale{1.07}\plottwo{hydro_sum7_sum98.eps}{nooscill_sum7_sum98_Xa_A.eps}
\epsscale{1.07}\plottwo{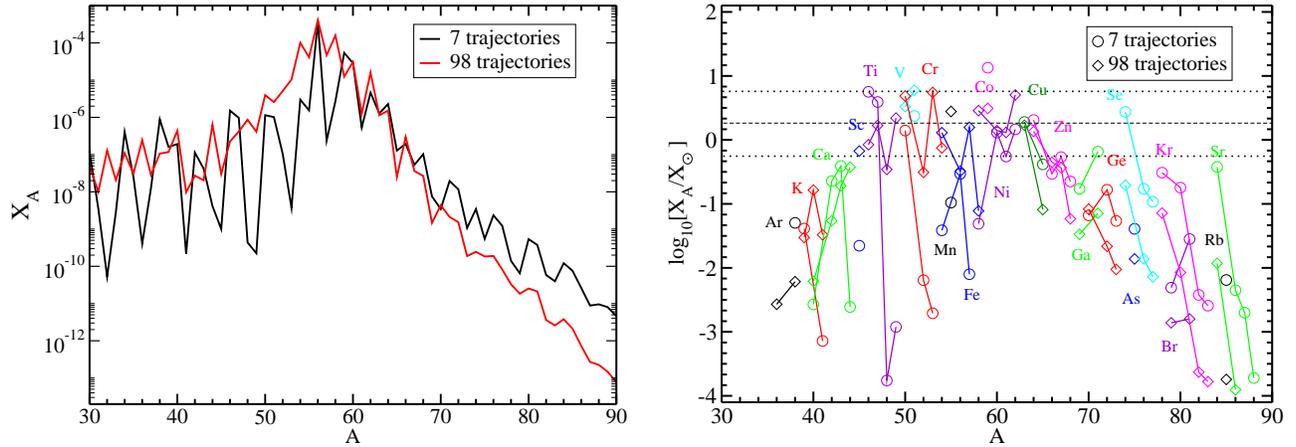}{f2b.eps}
\caption{{\em Left:} Mass fractions $X_{A}$ of the ejecta as a function of mass number $A$  comparing the cases
for 7 representative trajectories and for all the 98 trajectories.
{\em Right:} Comparison of the 
isotopic ejecta mass fractions ($X_{A}$) relative to the solar ones
 ($X_{\odot}$). The horizontal upper dotted line passes through the most
 overproduced isotopes ($^{51}$V, $^{53}$Cr, and $^{62}$Ni) in the 98 trajectory case, 
and the horizontal lower dotted line lies a factor 
of ten below the level of the upper line. The dashed line represents the median value. 
Our 7 selected trajectories reproduce the case with the 98 trajectories 
satisfactorily well only for certain values of $A$ (e.g., $58\leq A \leq 69$).
}
\label{fig:hydro_nucleosynth}
\end{figure*}
%--------------------------------------------

Figure~\ref{fig:hydro_nucleosynth} shows the nucleosynthesis mass fractions,
without taking into account neutrino oscillations, for the 98
trajectories and for the 7 trajectories after mass integration over the
ejecta mass-shell range as given by Eqs.~(\ref{eq:X98_A}) and (\ref{eq:X_A}), respectively.  
In the left panel, the mass fractions $X_{A}$ obtained for all of the 98 available
$\nu$-driven wind trajectories are compared to the ones obtained for
the 7 selected trajectories.  The right panel of
Fig.~\ref{fig:hydro_nucleosynth} shows the isotopic mass fractions $X_{A}$ 
relative to the solar ones $X_{\odot}$~\citep[][i.e., the production
factors]{Lodders03}  for the 98 available $\nu$-driven wind
trajectories and for the 7 representative ones as functions of $A$.
The dotted horizontal lines represent a ``normalization band.'' The
isotopes which fall into this band are considered to be the main nucleosynthetic products 
from the neutrino-driven wind phase of our fiducial ECSN model that could contribute to galactic chemical evolution. 
The upper dotted line passes through the most overproduced elements
($^{51}$V, $^{53}$Cr, and $^{62}$Ni), and the lower dotted line lies 
a factor of ten below that. The middle dashed line represents the median value. 

We find that the nucleosynthesis yields of the 7 trajectories reproduce 
those obtained from all the 98 trajectories only very approximately because
of the coarse time resolution of the wind history. Nevertheless,
this will be qualitatively sufficient to discuss the
effects of neutrino oscillations on the nucleosynthesis conditions.
The right panel of Fig.~~\ref{fig:hydro_nucleosynth} shows little production of isotopes
with $A > 65$ in the 98 trajectory case as well as in the 7 trajectory case. 
This is a consequence of a weak
$\nu$\textit{p}-process\footnote{In Table 1, we show the asymptotic values (indicated by subscript ``a'') 
of the electron fraction $Y_{e,\mathrm{a}}$ for our 7 representative trajectories. Notice that 
since $Y_{e,\mathrm{a}}>0.5$ for all the considered cases, the $\nu$\textit{p}-process 
may be enabled.} 
in this supernova environment because of the absence of a dense
outer stellar envelope in ECSNe, which is crucial for an efficient
$\nu$\textit{p}-process \citep{Wanajo2011b}. Many of the iron-group and
light trans-iron isotopes still lie on the normalization band, but the
greatest production factors (for $^{51}$V, $^{53}$Cr, and $^{62}$Ni in the 98 trajectory case) are
below 10. For example, the production factor of $^{62}$Ni is several times smaller 
than the corresponding one in the early
($\lesssim 400$~ms) convective ejecta, which are absent in 1D but found in
the 2D counterpart of the ECSN explosion model \citep{Wanajo2011aaa, Wanajo2013}. It appears,
therefore, that the nucleosynthetic contribution of the $\nu$-driven wind to
the Galactic chemical evolution is unimportant. It should be noted,
however, that the effects of nucleon potential corrections might 
alter the $Y_{e}$ history; thus the wind contribution
could be more important for nucleosynthesis than found here.

%%%%%%%%%%%%%%%%%%%%%%%%%%%%%%%%%%%%%%%%%%%%%%%%%%%%%%%%%%%%%%%%%%%%%%%%%%%%%%%%%%%%%%%%%%%%%%%%%%%%%%%%
\section{Reference neutrino signal and flavor evolution equations} \label{sec:neutrinos}
%%%%%%%%%%%%%%%%%%%%%%%%%%%%%%%%%%%%%%%%%%%%%%%%%%%%%%%%%%%%%%%%%%%%%%%%%%%%%%%%%%%%%%%%%%%%%%%%%%%%%%%%
At radius $r > R_\nu$, the unoscillated spectral number fluxes for each flavor
$\nu$ ($\nu = \nu_e, \bar{\nu}_{e}, \nu_x, \bar{\nu}_x$ with $x = \mu$ or~$\tau$) can be approximated by
%.............................................................
\begin{equation}
F_{\nu}(E) \approx \frac{L_{\nu}}{4 \pi r^2}\,\frac{f_{\nu}(E)}{\langle E_{\nu} \rangle} \ ,
\label{flux_nu}
\end{equation}
%.............................................................
where $L_{\nu}$ is the luminosity for the flavor $\nu$ and $\langle
E_{\nu} \rangle$ the mean spectral energy\footnote{In Eq.~(\ref{flux_nu}), general relativistic redshift corrections, 
which depend on $r$, as well as a ``flux factor'' accounting for nonradial  
neutrino momenta close to the neutrinosphere, are ignored.}. The neutrino
spectrum $f_{\nu}(E)$ is well reproduced by a combined power-law and exponential 
fit~\citep{Keil:2002in,Tamborra:2012ac}:
\begin{equation}
f_{\nu}(E)=\xi_\nu\left(\frac{E}{\langle E_{\nu} \rangle}\right)^{\alpha_\nu} e^{-(\alpha_\nu+1) 
E/\langle E_{\nu} \rangle}\ ,
\label{eq:f_E}
\end{equation}
being the parameter $\alpha_\nu$ defined by $\langle E_{\nu}^2
\rangle/\langle E_{\nu} \rangle^2 =
(2+\alpha_\nu)/(1+\alpha_\nu)$ and $\xi_\nu$ a normalization
factor such that $\int f_{\nu}(E) \,\ud E=1$. 

In order to incorporate neutrino oscillations
in our nucleosynthesis computations, we consider the 7
selected postbounce times $t_0$ as representative of the changing wind conditions
during the proto-neutron star cooling phase 
(note the partial overlap with data from the simulation by \citealp{Huedepohletal10} 
used for the analysis in~\citealp{Tamborraetal12}).
In Table~1 we list the neutrinosphere radius $R_\nu$ (assumed to be equal for all flavors), 
the luminosity $L_{\nu}$, the mean energy $\langle E_{\nu} \rangle$, and the
fit exponent $\alpha_\nu$  for each neutrino flavor and for the seven representative 
wind trajectories. 

In what follows, we neglect oscillations driven by the smallest mass difference between the active 
flavors, $\delta m_{\rm sol}$, and 
focus on neutrino oscillations in the active sector driven by the largest mass difference between
$\nu_e$ and $\nu_x$, $\delta m_{\rm atm}$, and by the mixing angle $\theta_{13}$. 
The reduction to two effective active flavors is justified, since oscillations driven by the 
solar parameters tend to
take place at a  radius larger than the one at which oscillations driven by $\delta m_{\rm atm}^2$ occur. Flavor
oscillations driven by the solar parameters are, therefore, unlikely 
to affect SN nucleosynthesis (see~\citealp{Dasgupta:2007ws,Fogli:2008fj,Dasgupta:2010cd} for details). 
Concerning active-sterile oscillations, 
we assume the mixing only of the electron neutrino flavor with a light sterile state for simplicity.
Overall,  we discuss a  2-flavor scenario (2 active flavors, $\nu_e$ and $\nu_x$) as well as a  
3-flavor one  (2 active+1 sterile flavors, $\nu_e$, $\nu_x$ and $\nu_s$). 
 
If interpreted in terms of sterile neutrinos $\nu_s$,
the reactor antineutrino anomaly requires a mass difference in the eV range, 
and cosmological hot dark matter limits imply that the sterile state
would have to be heavier than the active flavors~\citep{Abazajianetal12}. We here adopt the 
following mass splittings~\citep{Capozzi2013, Mention:2011rk}:
%...........................
\begin{eqnarray}
\delta m^2_{\rm atm} =-2.35\times 10^{-3}\mathrm{\ eV}^2\ \ {\mathrm{and}}\ \ 
\delta m^2_{\rm s} = 2.35 \mathrm{\ eV}^2\ ,
\end{eqnarray}
%...........................
with $\delta m^2_{\rm atm}$ being the squared mass difference between 
the neutrino mass eigenstates $\nu_3$ and the remaining two $\nu_{1,2}$~\citep{Fogli:2005cq} and 
$\delta m^2_{\rm s}$ the squared mass difference between 
the neutrino mass eigenstate $\nu_4$ and $\nu_{1}$,  chosen to be representative of reactor-inspired
values. 
We assume normal hierarchy
for the sterile mass-squared difference, namely $\delta m_{\rm s}^2 > 0$ 
(i.e., the neutrino mass eigenstate $\nu_4$ is heavier than the other 
mass eigenstates associated to the active neutrino flavors) and
inverted mass hierarchy for the atmospheric difference, $\delta m_{\rm
atm}^2 < 0$ (meaning that the neutrino mass eigenstate $\nu_3$ is lighter than 
$\nu_{1,2}$, see~\citealp{Fogli:2005cq}).  Note that current global fits of 
short-baseline neutrino experiments estimate 
$0.82 \le \delta m^2_{\rm s} \le 2.19$~eV$^2$ at 3$\sigma$ of confidence level~\citep{Giunti2013PhRvD}, 
which is lower than our adopted reference value~\citep{Mention:2011rk}.  
Our conservative choice favors a comparison with previous results discussed 
in~\cite{Tamborraetal12} besides not qualitatively changing our conclusions. 
We choose to scan only the inverted hierarchy scenario in the active sector (i.e., $\delta m_{\rm
atm}^2 < 0$), since this is the case where the largest impact due to collective flavor 
oscillations on nucleosynthesis is expected~\citep{Hannestad:2006nj,Fogli:2007bk,Fogli:2008pt,Dasgupta:2010cd}. 
The associated ``high'' (H) and sterile (S) vacuum oscillation frequencies are then
%........................................................
\begin{eqnarray}
\omega_{\mathrm H} = \frac{\delta m^2_{\mathrm{atm}}}{2E}\ \ {\mathrm{and}}\ \
\omega_{\mathrm S} = \frac{\delta m^2_{\mathrm s}}{2E}\ ,
\label{oscill_freq}
\end{eqnarray}
%.........................................................
with $E$ being the neutrino energy.
For the mixing angles we use~\citep{Capozzi2013, Mention:2011rk}
%....................................
\begin{eqnarray}
\label{theta13}
\sin^2 2 \theta_{14}= 10^{-1}\ \ {\mathrm{and}}\ \ \sin^2\theta_{13}=2 \times 10^{-2}\ .
\end{eqnarray}
%........................................

We treat neutrino oscillations in terms of matrices of
neutrino densities $\rho_E$ for each energy mode  $E$.
The diagonal elements of the density matrices are related to the neutrino densities, while the off-diagonal
ones encode phase information.
 The radial flavor evolution of the
neutrino flux is given by the ``Schr\"odinger equations,''
%....................................................
\begin{eqnarray}\label{eq:eom1}
\mathit{i} \partial_r\rho_E=[{\textsf H}_{E},\rho_{E}] \ \
{\mathrm {and}}\ \ 
\mathit{i}\partial_r\bar\rho_E=[\bar{\textsf H}_{E},\bar\rho_{E}]\,,
\end{eqnarray}
%....................................................
where an overbar refers to antineutrinos and sans-serif letters
denote $3{\times}3$ matrices in the ($\nu_e, \nu_x, \nu_s$) flavor space. 
The initial conditions for the density matrices are
$\rho_{E}=\mathrm{diag}(n_{\nu_e},n_{\nu_x},0)$ and $\bar\rho_{E} =
\mathrm{diag}(n_{\bar{\nu}_e},n_{\bar{\nu}_x},0)$, i.e., we assume that sterile
neutrinos are generated by flavor oscillations. The Hamiltonian matrix consists 
of the vacuum, matter and neutrino self-interaction terms: 
%....................................................
\begin{equation}
{\textsf H}_{E}= {\textsf H}^{\mathrm{vac}}_{E}+{\textsf H}^{\mathrm m}_{E}+{\textsf H}^{\nu\nu}_{E}\ .
\label{eq:ham}
\end{equation}
%....................................................
In the flavor basis, the vacuum term, 
%....................................................
\begin{equation}
{\textsf H}^{\mathrm{vac}}_{E} = {\textsf U}\,\mathrm{diag}\left(-\frac{\omega_{\mathrm H}}{2},+\frac{\omega_{\mathrm H}}{2},\omega_{\mathrm S}\right) {\textsf U}^{\dagger}\ ,
\end{equation}
%....................................................
is a function of the mass-squared differences 
(with ${\textsf U}$ being the 
unitary matrix transforming between the mass and the interaction basis) 
and of the mixing angles.
The matter term spanned by $(\nu_e,\nu_x,\nu_s)$ is in the flavor basis
%....................................................
\begin{eqnarray}
\label{lambda}
{\textsf H}^{\mathrm m} &=& \sqrt{2}G_{\rm F}\;
{\mathrm{diag}}(N_{e}-\frac{N_{n}}{2},-\frac{N_{n}}{2},0)\;,
\end{eqnarray}
%....................................................
with  $N_{e}$  the net electron number density 
and $N_{n}$ the neutron density.
Using Eq.~(1), the matter term becomes
\begin{eqnarray}
{\sf H}^{\rm m} &=& \sqrt{2}G_{\rm F} N_b \;
{\rm diag}\left(\frac{3}{2} Y_e -\frac{1}{2},\frac{1}{2} Y_e -\frac{1}{2},0\right)\;,
\label{HmYe}
\end{eqnarray}
being $N_b$ the baryon density. Note that the matter potential can be 
positive or negative and for $Y_e>1/3$ ($Y_e < 1/3$) a $\nu_e$-$\nu_s$ ($\bar{\nu}_e$-$\bar{\nu}_s$) 
MSW resonance can occur~\citep{wolf,Nunokawa:1997ct,McLaughlinetal99,Fetter:2000kf}. 
Because of Eq.~(\ref{HmYe}), neutrinos feel a different matter potential as $Y_e$ changes
and, at the same time,  $Y_e$ is affected by neutrino
oscillations via Eq.~(\ref{Yeeq}). 

The $\sf{H}^{\nu\nu}$ term describes $\nu$-$\nu$ interactions and vanishes for
all elements involving sterile
neutrinos~\citep{Sigl:1992fn}, i.e., ${\textsf H}^{\nu\nu}_{es}={\textsf
H}^{\nu\nu}_{xs}={\textsf H}^{\nu\nu}_{ss}=0$ 
(i.e., the only non-vanishing
off-diagonal element of the 3$\times$3 matrix is ${\textsf
H}^{\nu\nu}_{ex}$). In
the treatment of $\nu$-$\nu$ interactions,  we
assume the so-called ``single-angle approximation'' for the sake of simplicity, i.e.,  we assume that all neutrinos
feel the same average neutrino-neutrino refractive
effect~\citep{Duan:2006an,Fogli:2007bk,Duan:2010bg}.  
We will discuss in the following
the limits of such approximation (see Sect.~\ref{sec:discussion}).

In what follows, we explore the impact of  active-active and 
active-sterile neutrino conversions 
on the nucleosynthesis conditions and nucleosynthetic yields
for the 7 representative trajectories corresponding to 
postbounce times $t_{0}$. We distinguish two scenarios:
\begin{enumerate}
\item ``Active'' case, referring to neutrino oscillations in the active sector (2 active states).
\item ``Sterile'' case, meaning neutrino oscillations in the active and sterile sectors (2 active states + 1 sterile state).
\end{enumerate}
The coupled equations of the neutrino flavor evolution (Eqs.~\ref{eq:eom1}) were
discretized in the energy range $1$--$60$~MeV and solved by numerical integration 
together with Eq.~(\ref{Yeeq}) at each selected $t_0$~\footnote{Note that, for simplicity, in our computations we consider the 
effects of energy-dependent features of the oscillated neutrino spectra on the
$Y_e$ evolution in an integral sense by adopting 
neutrino  spectral quantities averaged over energy in Eqs.~(\ref{sigmae}, \ref{sigmaanue}).}. 
The initial conditions for the electron fraction and the neutrino spectral properties were assumed as given in 
Table 1.

%--------------------------------------------
\begin{figure*}
%\plottwo{IR_0_5s_r_L_E.eps}{IR_TOT_r_Ye_0_5s.eps}
\plottwo{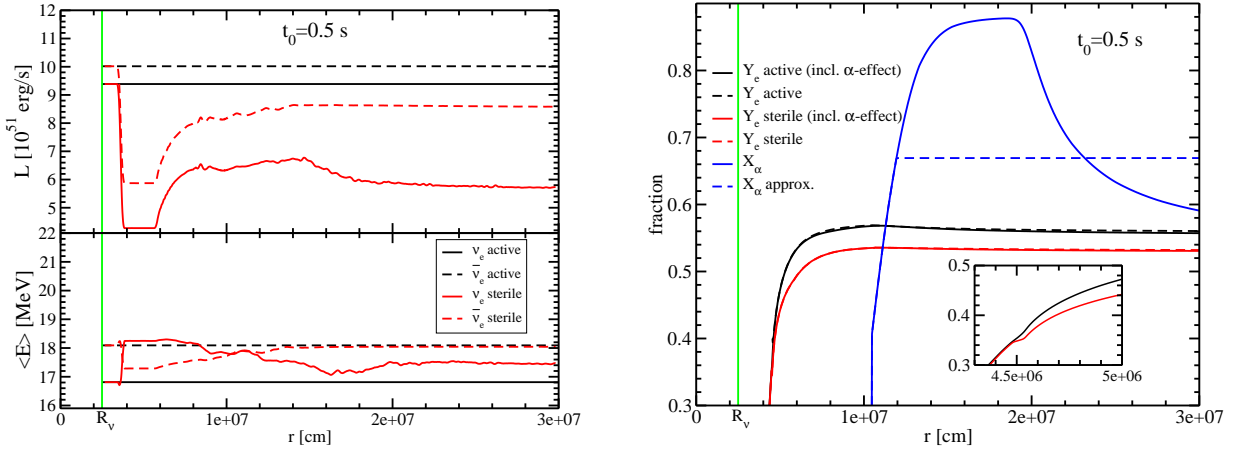}{f3b.eps}
\caption{{\em Left:} Electron neutrino and antineutrino luminosities
($L_{\nu_e}$ and $L_{\bar{\nu}_e}$) in units of $10^{51}$~erg$/$s (upper panel) 
and mean energies ($E_{\nu_e}$ and $E_{\bar{\nu}_e}$, lower panel) 
as functions of the distance ($r$) from the center of
the PNS at $t_{0}=0.5$ s postbounce. (The solid red lines 
are computed as running averages over $\Delta r \simeq 3.5 \times 10^5$ cm.) 
In the active case the luminosities and mean energies
of both $\nu_e$ and $\bar{\nu}_e$ are constant for $r \ge R_\nu$, which implies that the active 
case does not show any significant variations compared to the case 
without $\nu$ oscillations for the studied ECSN progenitor. 
In the sterile case, the inner active-sterile MSW resonance occurs for $\nu$ and $\bar{\nu}$ 
at $r \simeq 4 \times 10^6$~cm. Visible modifications of the neutrino spectral properties due to neutrino self-interactions
occur at $6 \times 10^6$~cm, while the outer MSW resonance occurs at about $1.4 \times 10^7$~cm.
{\em Right:} Electron fraction $Y_{e}$ and $\alpha$ mass fraction $X_\alpha$ as functions of 
distance $r$ from the center of the PNS at $t_{0}=0.5$~s.
In the active scenario 
neutrino oscillations negligibly affect $Y_e$ (the same as in the
no oscillations case which is not shown here).
The solid lines (``incl. $\alpha$-effect'' cases) refer to $Y_e$ obtained when
full network calculations are performed (the corresponding $X_\alpha$ is also shown with the solid blue line), 
while the dashed $Y_e$ lines refer to calculations corresponding to case (ii) in Sect.~\ref{sec:yeevo} 
(the corresponding $X_\alpha$ is also shown by a dashed blue line). 
The existence of the plateau in the $Y_e$ profile is shown in the inset of the 
right panel. 
The vertical line shows the neutrinosphere radius $R_\nu$.}
\label{fig:05s_L_E}
\end{figure*}

%--------------------------------------------
\begin{figure*}
%\plottwo{IR_f3a.eps}{IR_f3b.eps}
\plottwo{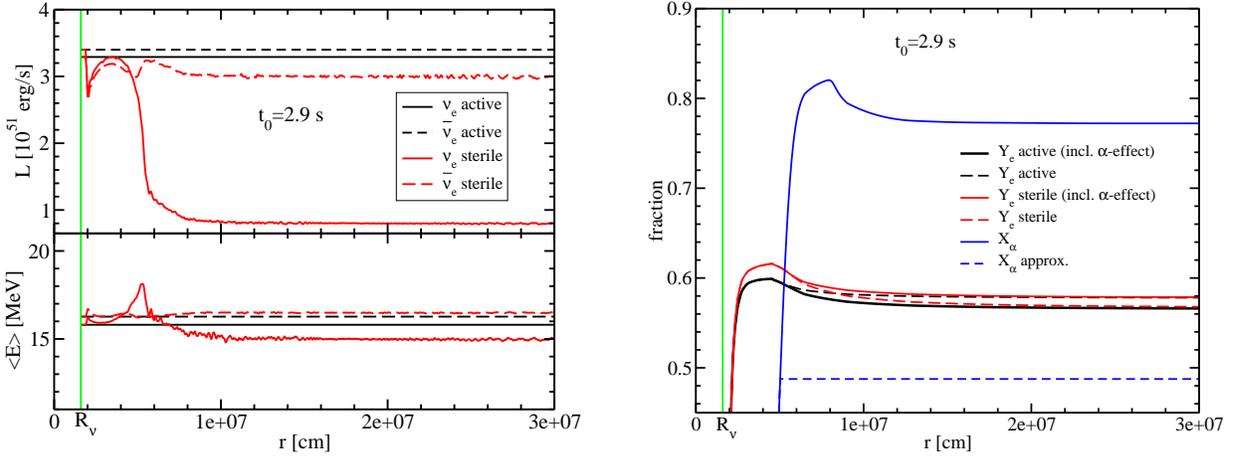}{f4b.eps}
\caption{Same as Fig.~\ref{fig:05s_L_E}, but at $t_{0}=2.9$~s.
In the sterile case, the inner active-sterile MSW resonance occurs for $\nu$ and $\bar{\nu}$ 
at $r \simeq 2 \times 10^6$~cm. The outer MSW resonance occurs at about $8 \times 10^6$~cm.
(The solid red lines are computed as running averages over $\Delta r \simeq 2.9 \times 10^5$ cm.)
}
\label{fig:29s_L_E}
\end{figure*}

%--------------------------------------------
\begin{figure*}
%\plottwo{IR_f4a.eps}{IR_f4b.eps}
\plottwo{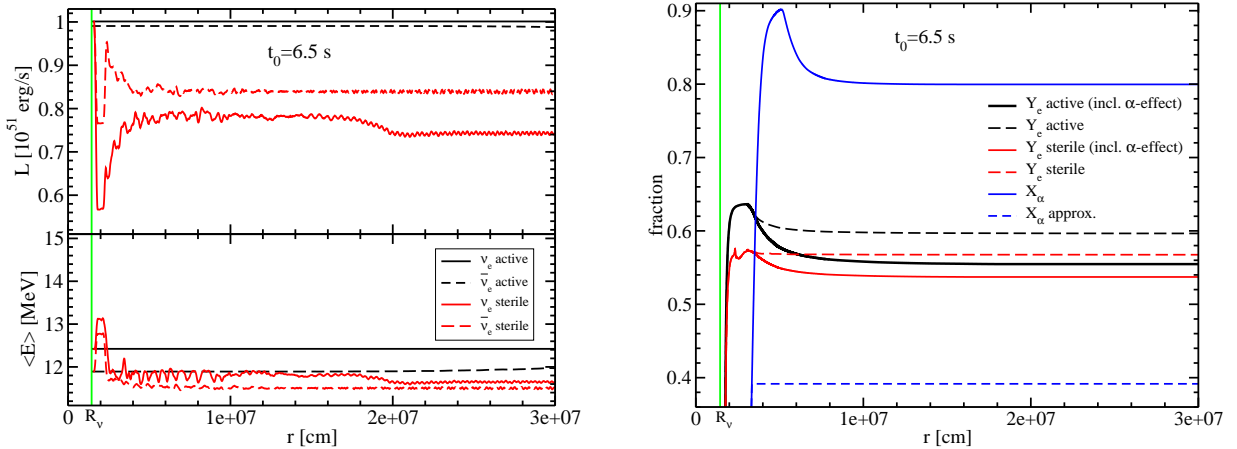}{f5b.eps}
\caption{{\em Left:} Same as Fig.~\ref{fig:05s_L_E}, but at $t_{0}=6.5$~s.
In the sterile case the inner MSW resonance occurs for $\nu$ and $\bar{\nu}$ at $r\simeq 1.8\times10^6$ cm. 
The outer $\nu_e$-$\nu_s$ MSW takes place at $r\simeq 2.5\times10^6$~cm.
(The red lines display running averages over $\Delta r \simeq 2.1 \times 10^5$ cm.) 
}
\label{fig:6_5s_r_L_E}
\end{figure*}
%....................................................

%%%%%%%%%%%%%%%%%%%%%%%%%%%%%%%%%%%%%%%%%%%%%%%%%%%%%%%%%%%%%%%%%%%%%%%%%%%%%%%%%%%%%%%%%%%%%%%%%%%%%%%%%%%%
\section{Neutrino oscillations in the neutrino-driven wind and feedback on the electron fraction} \label{sec:Ye_oscil}%
%%%%%%%%%%%%%%%%%%%%%%%%%%%%%%%%%%%%%%%%%%%%%%%%%%%%%%%%%%%%%%%%%%%%%%%%%%%%%%%%%%%%%%%%%%%%%%%%%%%%%%%%%%%%

In this section, we discuss the neutrino flavor oscillation physics during 
the neutrino-driven wind phase and the oscillation feedback on $Y_e$ for scenarios 1 and 2 
(see Sect.~\ref{sec:neutrinos}). After qualitatively describing 
the oscillation phenomenology, we will discuss in detail how the neutrino fluxes are affected by flavor oscillations 
at three representative times $t_0= 0.5, 2.9$ and $6.5$~s, 
representing the early, intermediate and late cooling phases, respectively.
We will focus on the impact of flavor oscillations on  $Y_e$, neglecting the $\alpha$-effect for 
the sake of simplicity 
(i.e., $X_\alpha$ is assumed to be as in case (ii) in Sect.~3); 
The role of the $\alpha$-effect on the electron fraction and 
its interplay with neutrino oscillations will be described in Sect.~\ref{sec:Ye_Xa}.

\subsection{Neutrino oscillation phenomenology}

In the presence of only active neutrinos, the MSW resonance due to the atmospheric mass difference occurs at  radii
much larger than the ones considered here ($r \lesssim 3 \times 10^7$~cm), where $Y_e$ has already reached its 
asymptotic value, and therefore the electron fraction is not affected. Because of $\nu$-$\nu$ interactions, 
 multiple spectral splits should occur in inverted 
hierarchy for the initial conditions of neutrinos and antineutrinos of the studied ECSN
(i.e., $L_{\nu_e,\bar{\nu}_e}/\langle E_{\nu_e,\bar{\nu}_e} \rangle - L_{\nu_x}/\langle E_{\nu_x} \rangle < 0$,~\citealp{Fogli2009JCAP}).  
However, since the $\nu_e$ and $\bar{\nu}_e$ luminosities and mean energies are very similar 
to those of the heavy-lepton 
neutrinos, as shown in Table~1, and because of the total lepton-number conservation, 
we do not expect any appreciable variations  in the oscillated luminosities and mean energies 
(see \citealp{Fogli2009JCAP} for an
extended discussion). 
 
In the sterile scenario,  while active neutrinos propagate away from the SN core, they interact with the matter background and 
convert to sterile states
through MSW resonances in two different 
spatial regions (see also Appendix~\ref{sec:AppendixA}).  Close to the neutrinosphere, 
due to the steep growth of $Y_e$, and therefore of the matter potential via Eq.~(\ref{HmYe}), 
the inner active-sterile MSW  resonance occurs for both neutrinos and antineutrinos at about the same radius ($r_{\rm IR}$). 
At larger radii (located closer to the neutrinosphere
as the postbounce time increases), an outer active-sterile MSW resonance occurs and it mainly 
affects neutrinos.
Any modification of the neutrino energy spectra due to oscillations will affect the electron fraction via Eqs.~(\ref{Yeeq}), 
(\ref{lambdanue}) and (\ref{lambdaantinue}). 

At early postbounce times, the matter potential  felt by neutrinos close to $r_{\rm IR}$ is 
slightly less steep than the one felt by antineutrinos (see 
Appendix~\ref{sec:AppendixA} and left panel of Fig.~\ref{fig:TOT_r_Ye}), 
therefore the 
 adiabaticity of the $\bar{\nu}$ conversions is slightly decreased and the one
of $\nu$ slightly increased with a net  conversion probability for neutrinos a bit larger than for 
antineutrinos,  as pointed out by \cite{Nunokawa:1997ct}. 
This is particularly evident during the accretion phase as discussed in~\cite{Wuetal2013} and, 
for our purposes,  during  the early-cooling  phase ($t_0=0.5, 1$~s), where the difference in the conversion
probabilities of $\nu_e$ and $\bar{\nu}_e$ is responsible for a plateau in the $Y_e$ 
profile close to $r_{\rm IR}$ ~\citep{Wuetal2013}. 
As the postbounce time increases, the matter potential  felt by neutrinos close to 
$r_{\rm IR}$ becomes steeper (see Fig.~\ref{fig:TOT_r_Ye}, left panel), and therefore 
the ${\nu}_e \rightarrow {\nu}_s$ and $\bar{\nu}_e \rightarrow \bar{\nu}_s$ resonant conversions 
are expected to roughly have the same degree of adiabaticity, with a resultant small feedback effect on $Y_e$ 
(assuming that further flavor conversions due to $\nu$-$\nu$ interactions are negligible).

The outer active-sterile MSW resonance is generally more adiabatic than the inner one: It occurs where the matter potential 
is shallow and the effective mixing angle is larger. Therefore, $\nu_e$ are abundantly converted 
to $\nu_s$, lowering the wind $Y_e$ (via Eqs.~\ref{Yeeq}, \ref{lambdanue}, and \ref{lambdaantinue}). 

Besides neutrino interactions with matter, neutrino self-interactions affect the neutrino oscillated fluxes, and therefore
$Y_e$ (see Appendix~\ref{sec:AppendixA} for more details). As discussed in~\cite{Tamborraetal12},  $\nu_e \leftrightarrow \nu_x$ conversions,  
due to neutrino-neutrino interactions, partially repopulate the electron sector depleted by  $\nu_e \rightarrow \nu_s$ MSW conversions. 
The net effect  is that $\nu$-$\nu$
interactions favor the repopulation of the $\nu_e$ sector (because of
$\nu_x$-$\nu_e$ conversions) and partially counterbalance the effect of the
$\nu_e$-$\nu_s$ MSW resonances on the electron fraction.
The role played by neutrino self-interactions becomes more and more evident as
the time $t_0$ increases, since the matter background is lower.

\subsection{Results: Neutrino oscillations and feedback on the electron fraction} 

In order to quantitatively describe the impact of flavor oscillations on the $Y_e$ evolution as $t_0$ increases,  we 
select three representative postbounce times,  $t_0=$ 0.5 s, 2.9 s, and 6.5 s, and discuss the oscillation 
phenomenology in the active and sterile cases.

Figure~\ref{fig:05s_L_E} (left panel) shows the luminosities and mean energies
for $\nu_e$ and $\bar{\nu}_e$ as functions of the radius in the
active and sterile cases at $t_0 = 0.5$~s.  
As expected, in the active case, neutrino
oscillations do not visibly modify the mean energies and the luminosities in the 
radial regime where $Y_e$ is still evolving (i.e., $r \lesssim 2 \times 10^7$~cm).
To demonstrate the effect of neutrino oscillations  on $Y_{e}$, we plot $Y_e$ as a function of the radius at 
$t_0=0.5$~s in Fig.~\ref{fig:05s_L_E} (right panel). In the active case, 
the $Y_{e}$ evolution does not differ from the case without neutrino oscillations. 
In the sterile case, instead, the inner active-sterile MSW resonance occurs 
at $r_{\rm IR} \simeq 4 \times 10^6$~cm 
and leads to the formation of a small plateau in the $Y_e$ profile (see zoom in the right panel of Fig.~3). 
In fact, the inner resonance is responsible for a $\nu_e \rightarrow \nu_s$ 
conversion probability larger than the $\bar{\nu}_e \rightarrow \bar{\nu}_s$ one, 
as expected (Fig.~\ref{fig:05s_L_E}, left panel). Such active-sterile flavor conversion modifies the 
$\nu_e$ and $\bar{\nu}_e$ energy spectra,  introducing non-zero off-diagonal terms in the neutrino density matrices. Neutrino self-interactions
are therefore triggered at about $6 \times 10^6$~cm. The outer active-sterile MSW resonance
occurs at  $r_{\rm OR} \simeq 1.4 \times 10^7$~cm, converting $\nu_e$ to  $\nu_s$. 
The corresponding electron fraction (Fig.~\ref{fig:05s_L_E}, right panel) 
shows a  very small plateau in correspondence
of $r_{\rm IR}$ and it remains lower than in the active case due to active-sterile flavor conversions.

Figure~\ref{fig:29s_L_E}, analogously to Fig.~\ref{fig:05s_L_E}, shows the luminosities and mean energies
for $\nu_e$ and $\bar{\nu}_e$ as functions of the radius at $t_0 = 2.9$~s (left panel) and the corresponding 
electron fraction (right panel).  
In the active case, neutrino
oscillations do not visibly modify the neutrino spectral properties in the 
radial regime where $Y_e$ is still evolving, as already discussed at $t_0 = 0.5$~s.
In the sterile case, the inner active-sterile MSW resonance occurs at $r_{\rm IR} \simeq 2 \times 10^6$~cm.
As discussed in Appendix~\ref{sec:AppendixA}, the instability induced by the inner MSW resonance and the fact that the matter 
potential is lower than at earlier postbounce times trigger neutrino self-interactions converting 
slightly more $\bar{\nu}_e$ than $\nu_e$, contrarily to what is expected. The outer active-sterile MSW resonance
occurs at about $r_\mathrm{OR} \simeq 5 \times 10^6$~cm converting a large number of $\nu_e$ 
to $\nu_s$. Correspondingly,
the $Y_e$ profile (Fig.~\ref{fig:29s_L_E}, right panel) is higher than the active one close to the neutrinosphere (because more $\bar{\nu}_e$
are converted to sterile states than $\nu_e$). The depletion of the $\nu_e$ flux due to the outer MSW resonance is responsible for
lowering the electron fraction below the  active one 
(compare the black dashed line to the red dashed line).

Figure~\ref{fig:6_5s_r_L_E}  shows the radial evolution of the $\nu_e$ and $\bar{\nu}_e$ spectral properties
and the corresponding $Y_e$ profile at $t_{0}=6.5\,\mathrm{s}$.  
In this case as well, active neutrino oscillations do not change the values of 
the luminosities and mean energies, and therefore $Y_e$ does not change compared to the case without oscillations.
In the sterile case, instead, the inner MSW resonance  already occurs  at $r \simeq 1.8 \times10^{6}$~cm for $\nu$ and $\bar{\nu}$, triggering
at the same time neutrino collective oscillations, while the outer MSW 
resonance takes place at $r_{\mathrm{OR}} \simeq 2.5 \times 10^6$~cm (see Appendix~\ref{sec:AppendixA} for more details).
Finally, flavor conversions among the active flavors slightly modify the luminosity and mean energy of 
$\nu_e$ at $r\simeq2\times10^7$ cm, without affecting the survival probabilities (see Fig.~12). 
The MSW resonances together with $\nu$-$\nu$ interactions significantly reduce 
the $\nu_e$ number flux (i.e., $L_{\nu_e}/\langle E_{\nu_e} \rangle$) 
compared to the $\bar \nu_e$ number flux. 
This means that  a more neutron-rich environment (i.e., a lower $Y_e$) is favored compared to the active case
(see Fig.~\ref{fig:6_5s_r_L_E}, right panel). 

%%%%%

%%%%%%%%%%%%%%%%%%%%%%%%%%%%%%%%%%%%%%%%%%%%%%%%%%%%%%%%%%%%%%%%%%%%%%%%%%%%%%%%%%%%%%%%%%%%%%%%%%%%%%%%%%%%
\section{Interplay of neutrino oscillations and $\alpha$-effect on the electron fraction} \label{sec:Ye_Xa}%
%%%%%%%%%%%%%%%%%%%%%%%%%%%%%%%%%%%%%%%%%%%%%%%%%%%%%%%%%%%%%%%%%%%%%%%%%%%%%%%%%%%%%%%%%%%%%%%%%%%%%%%%%%%%

In this section, we discuss the evolution of $Y_e$ as a function of
radius at our selected postbounce times ($t_{0} =$ 0.5~s, 1~s, 2~s, 
2.9~s, 4.5~s, 6.5~s, and 7.5~s), for the scenarios 1 and 2 described in
Sect.~\ref{sec:neutrinos}, and with the two different assumptions made 
in Sect. 3 about the evolution of the mass fraction of 
$\alpha$ particles. These assumptions allow us to disentangle between 
the role played by neutrino oscillations and the $\alpha$-effect
in determining $Y_e$.  

The evolution of the electron fraction is not just influenced
by the $\nu_e$ and  $\bar\nu_e$ properties, 
which are affected by neutrino oscillations, as discussed in the 
previous section, but also by the presence of 
$\alpha$ particles (see Eqs.~\ref{Yeeq}, 5, and 6). 
Therefore, the whole $Y_e$ evolution 
is a complicated interplay between neutrino oscillations and the $\alpha$-effect. 
The outcome depends 
on the location of the region of active-sterile conversions relative to that of the
$\alpha$ particle formation.
For this reason, we choose to analyze the evolution of $Y_e$ in detail
at three representative postbounce times, $t_0=$ 0.5 s, 2.9 s, and 6.5 s. 

In Fig.~\ref{fig:05s_L_E} (right), we show the evolution of Y$_e$ at $t_{0}=0.5$ s, in
the active and sterile cases and 
with (``incl. $\alpha$-effect'' case) or without the inclusion of the $\alpha$-effect.
In this case, the formation of $\alpha$ particles does not play any significant role 
in determining Y$_e$, because the formation of $\alpha$ particles (solid blue line) occurs 
when $Y_e$ has almost reached its asymptotic value (compare the solid and dashed lines).\\
At intermediate and late postbounce times, the results of simulations with and without $\alpha$ particle formation
from free nucleons have to be distinguished, because the $\alpha$-effect associated
with the presence of large abundances of $\alpha$ particles has severe consequences
for the $Y_e$ evolution.
In Fig.~\ref{fig:29s_L_E} (right), we show the evolution of Y$_e$ at $t_{0}=2.9$ s,
analogously to Fig.~\ref{fig:05s_L_E} (right). In this case, the formation of 
$\alpha$ particles occurs when Y$_e$ is still evolving and it 
overlaps with the region where the outer MSW resonance
takes place (see Fig.~\ref{fig:29s_L_E}, left).

The results with the $\alpha$-effect 
(solid red and black lines in Fig.~\ref{fig:29s_L_E}, right) show a counterintuitive behavior.
While for active flavor oscillations the $\alpha$-effect 
drives $Y_e$ closer to 0.5 in the usual way (compare the black dashed and solid 
lines in Fig.~\ref{fig:29s_L_E}, right), the sterile neutrino case exhibits the opposite trend: 
In the presence of a higher abundance of $\alpha$ particles, i.e., despite the 
$\alpha$-effect, $Y_e$ remains higher and the evolution towards $Y_e=0.5$
is clearly damped (red solid line in comparison to red dashed line).
The formation of a larger abundance of $\alpha$ particles thus obviously 
reduces the influence of the active-sterile $\nu_e$-$\nu_s$ conversions on $Y_e$.
This astonishing result is a consequence of the fact that the conversion 
to sterile neutrinos occurs slightly outside (or overlaps with) the region
where the rapid recombination of neutrons and protons to $\alpha$ particles
takes place. In such a situation the influence of the $\nu_e$-$\nu_s$ conversion
on the $Y_e$ evolution is diminished by the lower number fractions of 
free neutrons and protons, which lead to a lower rate of change of $Y_e$
according to Eq.~(4). Instead of undergoing reactions with $\nu_e$ or 
$\bar{\nu}_e$, the majority of free nucleons react to form $\alpha$ particles 
as the wind expands away from the $\nu_e$-$\nu_s$ conversion radius.

The influence of $\alpha$ particle formation manifests itself differently in 
the late wind evolution, where $\nu_e$ conversions to sterile neutrinos take
place closer to the neutrinosphere and, in particular, at a radius which is
smaller than the one at which nucleon recombination begins 
to raise the $\alpha$ abundance.

In Fig.~\ref{fig:6_5s_r_L_E} (right), we display the evolution of 
the electron fraction $Y_e$ at $t_0=6.5$ s, in the active and sterile cases,
in analogy to Fig.~\ref{fig:29s_L_E} (right). 

In the sterile case, $Y_e$ is lower than 
in the active case already very close to the neutrinosphere where the matter is still in 
NSE (and thus no $\alpha$ particles are present).
The dashed lines are again calculated without the $\alpha-$effect,
while the solid lines include the $\alpha$-effect.

When the $\alpha-$effect is included, 
the value of $Y_e$ is, as expected, pushed towards 0.5 in both active
(black solid line) and sterile cases (red solid line).  
We notice that at $t_0=6.5$ s, differently from $t_0=2.9$ s,
neutrino oscillations, in particular both the inner and outer MSW resonances, 
take place \textit{before}
$\alpha$ particles start forming, and therefore they make the environment 
significantly less proton-rich ($Y_e$ is lowered) before the $\alpha-$effect takes place
and decreases $Y_e$ even further towards more symmetric conditions ($Y_e=0.5$)
in the usual way. 

\begin{figure*}
%\plottwo{IR_f5a.eps}{IR_f5b.eps}
\plottwo{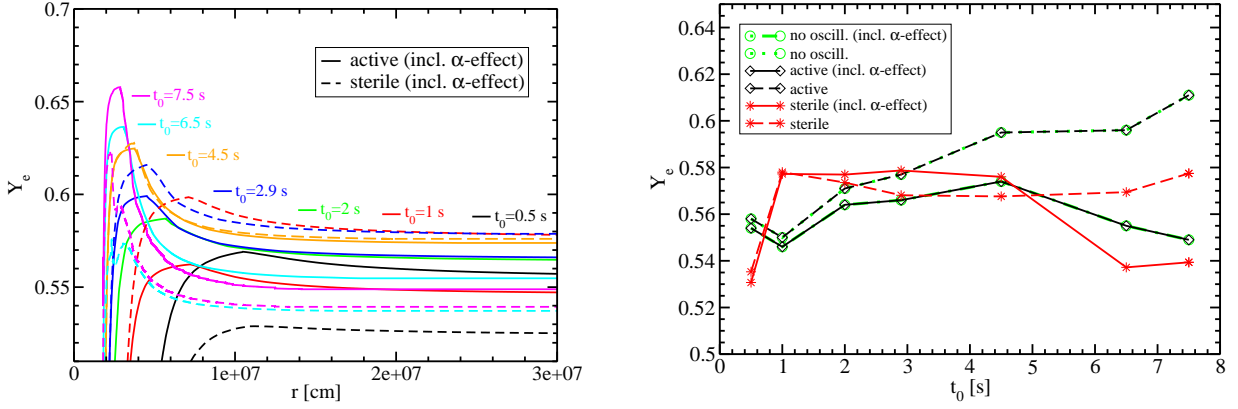}{f6b.eps}
\caption{{\em Left:} Electron fraction ($Y_{e}$) as a function of the distance $r$ from the center of the PNS 
at all considered postbounce times ($t_{\mathrm{0}}$), and in the active and sterile cases.
The $\alpha$-effect is included in all cases (``incl. $\alpha$-effect''). 
Because of the near equality of the neutrino luminosities and mean energies of neutrinos of 
all flavors, $Y_{e}$ in the active cases does not appreciably differ from the one obtained without neutrino oscillations.
{\em Right:} Asymptotic electron fractions ($Y_{e}$) as functions of postbounce time ($t_{0}$) 
in the active and sterile as well as no oscillations cases. 
The dashed lines refer to $Y_{e}$ calculated without the $\alpha$-effect, 
while the solid lines refer to $Y_{e}$
calculated with the full network. The $\alpha$-effect is stronger 
especially at late times ($t_{0}=6.5$ and 7.5 s) when the neutron star is more compact
and the neutrino luminosities are lower. The values in the cases 
without oscillations coincide with those in the active cases
and cannot be distinguished.}
\label{fig:TOT_r_Ye}
\end{figure*}

Figure ~\ref{fig:TOT_r_Ye} gives an overview of the interplay 
between neutrino oscillations and the $\alpha$-effect by showing
the evolution of the electron fraction $Y_{e}$ at all considered postbounce 
times $t_0$.
 
%--------------------------------------------
%--------------------------------------------

%--------------------------------------------
\begin{figure*}
%\plottwo{IR_f6a.eps}{IR_f6b.eps}
\plottwo{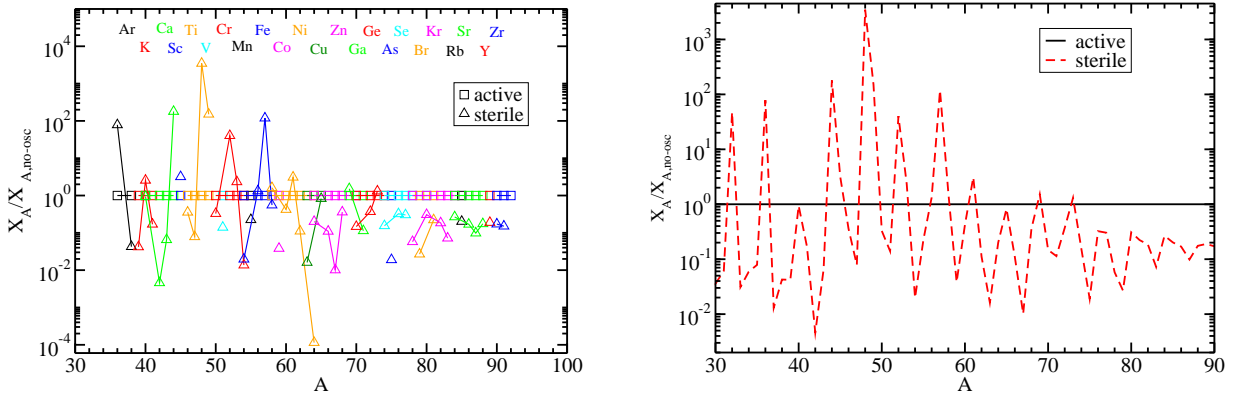}{f7b.eps}
\caption{{\em Left:} Isotopic mass fractions in the active and sterile cases
relative to those in the case without oscillations versus mass
number $A$ for all representative 7 trajectories. 
{\em Right:} Nucleosynthetic abundances in the active and
sterile cases relative to those without oscillations for all
the representative 7 trajectories. Since in our model active flavor 
oscillations do not change the neutrino properties and the wind $Y_e$ at any significant
level up to the radius of interest, the nucleosynthesis results are essentially 
identical for all the cases with active oscillations and no neutrino oscillations.}
\label{fig:oscill_abund}
\end{figure*}
%--------------------------------------------

Figure~\ref{fig:TOT_r_Ye} (left) shows $Y_{e}$ as a function of the distance
$r$ from the center of the PNS at different postbounce times $t_0$ 
in both the active and sterile cases and including the $\alpha$-effect. 

In Fig.~\ref{fig:TOT_r_Ye} (right) the asymptotic $Y_{e}$ values (namely, $Y_{e}$
at $r \simeq 3\times 10^7$~cm) are plotted as functions of the
postbounce time for each of the considered scenarios (active, sterile 
and no oscillations cases). 
Note that the values in the active case cannot be distinguished from those in the
no oscillations case, suggesting essentially negligible roles of the
active-active oscillations on the evolution of $Y_e$ (see discussion in Sect.~5).

Furthermore, in the active case, $Y_e$ is systematically pushed towards 0.5 by the
$\alpha$-effect, as we can see by comparing the black dashed line with the black solid one (``incl. $\alpha$-effect'' cases).
In the sterile case (red solid line), neutrino oscillations combined 
with the $\alpha$-effect lead to $Y_e$ being lower than 
in the active case (black solid line) at early postbounce times ($t_0=0.5$ s), higher than in the active case
at intermediate postbounce times ($t_0=1$ s, $2$ s, and $2.9$ s) and again lower 
than in the active case at late postbounce times  ($t_0=6.5$ s, and $7.5$ s).

In particular,  at late times, $Y_e$ in the sterile case and including the 
$\alpha$-effect becomes lower than $Y_e$  in the active case and 
lower than $Y_e$ in the case without full $\alpha$ recombination, because
both MSW $\nu_e$-$\nu_s$ conversions happen so close to the neutrinosphere
that the $\alpha$ particle formation at larger radii further enhances 
the $Y_e$-reduction associated with the presence of sterile neutrinos,
although $Y_e$ remains always higher than 0.5.
 
In summary, the $\alpha$-effect plays an important role in lowering
$Y_{e}$ especially at late times ($t_{0}=6.5$ s and $7.5$ s). This
is due to the higher entropy and the longer expansion timescale as a
result of the more compact PNS with the lower neutrino luminosities,
resulting in a delay of the $\alpha$ recombination relative to both the MSW 
$\nu_e$-$\nu_s$ conversions and to a longer duration of
the $\alpha$-effect (see also next section for more details). However,
although the $\alpha$-effect has a strong impact on $Y_e$ and therefore
on the element production, it plays only a sub-leading role for the neutrino
oscillations and no detectable modifications are expected for the
neutrino fluxes at the Earth.

Because of the leading role of the $\alpha$-effect compared to
oscillations on $Y_{e}$, especially at late times (see
Fig.~\ref{fig:TOT_r_Ye}, where $Y_{e}$ in the active and sterile cases 
including the $\alpha$-effect is fairly similar), we expect that the nucleosynthesis yields in the
presence of oscillations are not significantly different from the cases
where oscillations are not considered (see
Sect.~\ref{subsec:no_oscill_nucl}). This can be seen in
Fig.~\ref{fig:oscill_abund}, where we show the nucleosynthesis yields
obtained for the 7 representative trajectories in the active and
sterile cases relative to those without neutrino oscillations.  
In Fig.~\ref{fig:oscill_abund} (left) we notice that most of 
the isotopic mass fraction ratios in the sterile case relative 
to the no oscillation case are lower than 2, with the exception
of some isotopes (with $A<60$) which have enhanced production factors.

The most abundantly produced isotope in the relative comparison is 
$^{49}$Ti (X$^{^{49}\mathrm{Ti}}_{\mathrm{sterile}}/X^{^{49}\mathrm{Ti}}_{\mathrm{no-oscill.}}\simeq3.57\cdot10^{3}$).
This overproduction of the $^{49}$Ti isotope in the sterile case
compared to the case without oscillations, however, is still too small to have 
any significant impact on the production factor of this isotope (see Fig.~2, right). 
From Fig.~\ref{fig:oscill_abund}, it is also clear that, in the sterile case,
there is less production of heavy elements (e.g., $A\geq 70$)
than in the case without oscillations.  

For all the reasons above, one can conclude from Figs.~\ref{fig:TOT_r_Ye} and \ref{fig:oscill_abund} 
that neither active neutrino oscillations nor a fourth sterile neutrino family
can alter the nucleosynthesis-relevant conditions, nor can they create a
neutron-rich site ($Y_e < 0.5$) to activate the r-process in
the adopted ECSN model (without nucleon potential corrections; see
Sect.~7).

%--------------------------------------------
\begin{figure*}
%\plottwo{Toy1_2_9s_r_L_E.eps}{Toy1_r_Ye_2_9s.eps}
\plottwo{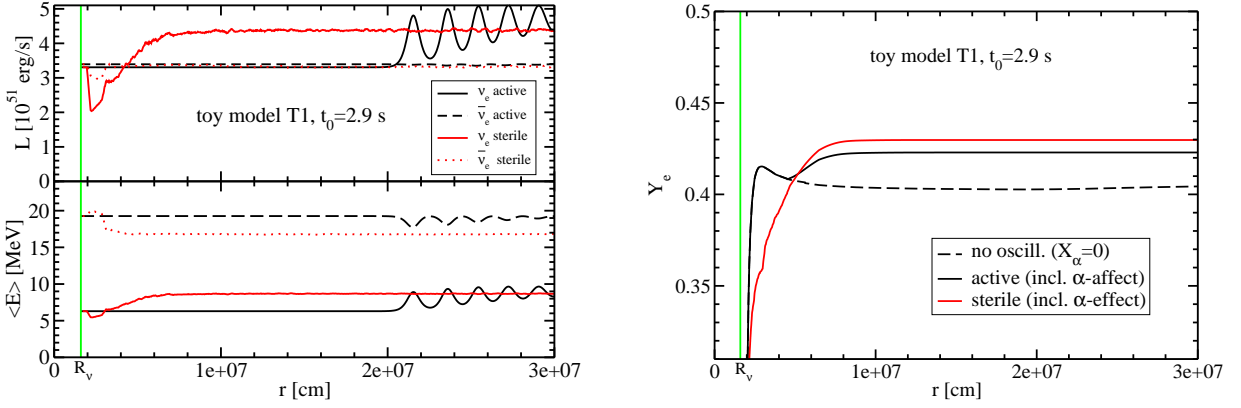}{f8b.eps}
\caption{{\em Left:} Electron neutrino and antineutrino luminosities ($L_{\nu_e}$ and $L_{\bar{\nu}_e}$ in units of $10^{51}$~erg$/$s, upper panel) 
for toy model 1 (see Table 2 and text for details) as functions of the distance $r$
from the center of the PNS, at $t_{0}=2.9$~s, in the active and sterile cases. Lower panel: Similar to the upper panel, 
but for the mean energies $\langle E_{\nu_e} \rangle$  and $\langle E_{\bar{\nu}_e} \rangle$. 
(The red lines are running averages over $\Delta r \simeq 1.98\times 10^5$ cm.)
{\em Right:} Electron fraction $Y_{e}$ as function of the distance $r$ from the center of the 
PNS for our toy model at $t_{0}=2.9$~s (see text for details) in the case without neutrino oscillations and 
setting $X_{\alpha} = 0$ (``no oscill. ($X_{\alpha}=0$)'' case, dashed black line), 
in the case with flavor conversions of active neutrinos (solid black line), and in the 
case of active-sterile conversions (solid red line). Both of the last two cases were computed with $\alpha$
particle recombination.
Neutrino oscillations, jointly with the $\alpha$-effect, 
drive $Y_{e}$ towards 0.5, disfavoring  the r-process.}
\label{fig:Toy1_Ye_2.9s}
\end{figure*}
%--------------------------------------------------------------------------------

%--------------------------------------------
\begin{figure*}
%\plottwo{Toy1_6_5s_r_L_E.eps}{Toy1_r_Ye_6_5s.eps}
\plottwo{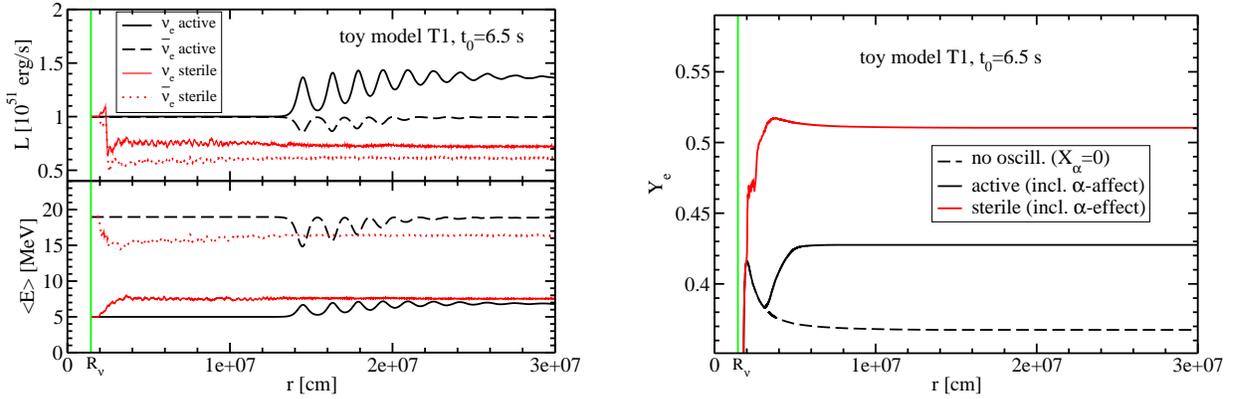}{f9b.eps}
\caption{Same as Fig.~\ref{fig:Toy1_Ye_2.9s}, but for toy model 1 at $t_{0}=6.5$ s (see text for details).
(The red lines here are running averages over $\Delta r \simeq 1.1\times 10^5$ cm.)}
\label{fig:Toy1_Ye_6.5s}
\end{figure*}
%--------------------------------------------------------------------------------

%--------------------------------------------
\begin{figure*}
%\plottwo{Toy3_2_9s_r_L_E.eps}{Toy3_r_Ye_2_9s.eps}
\plottwo{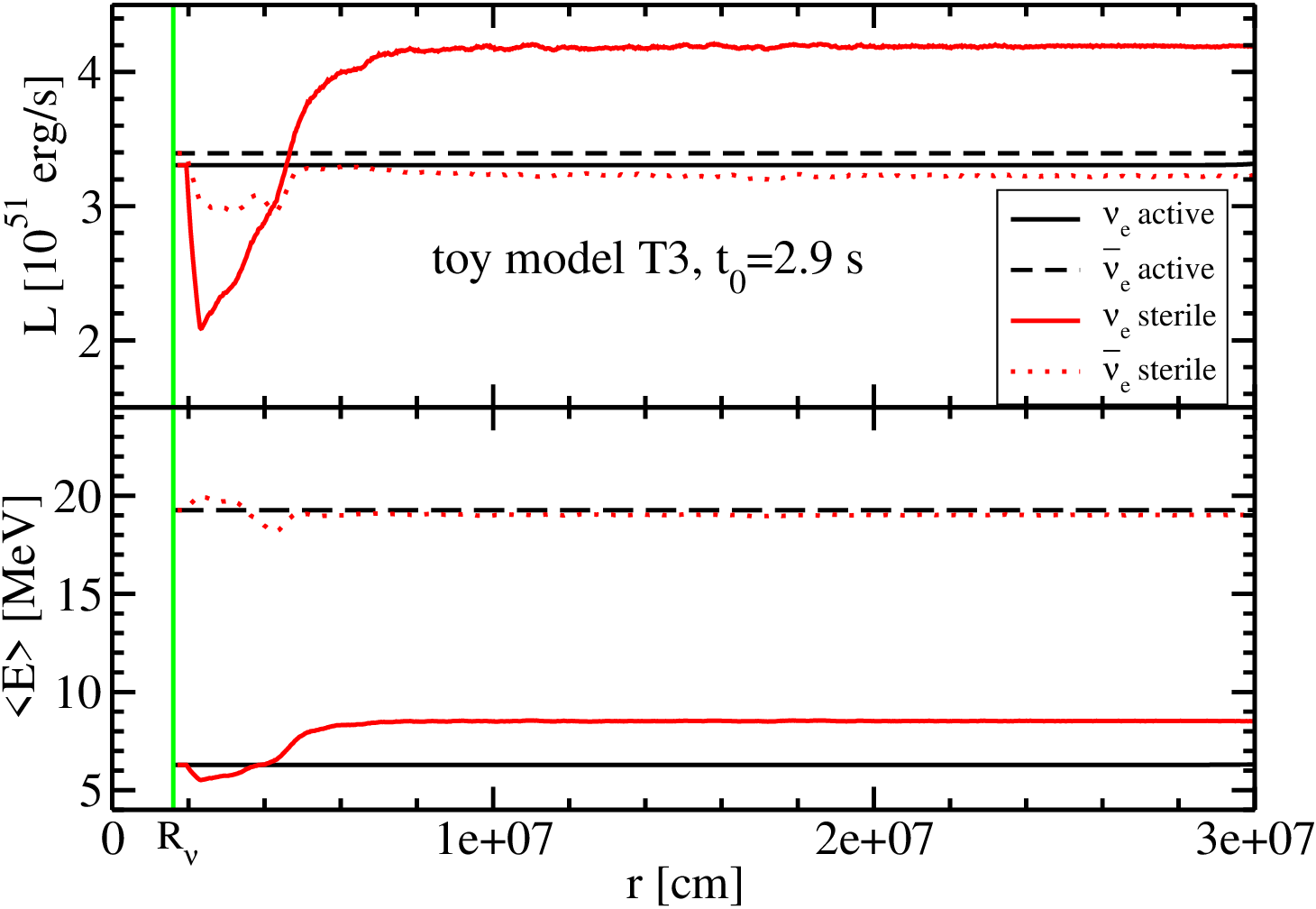}{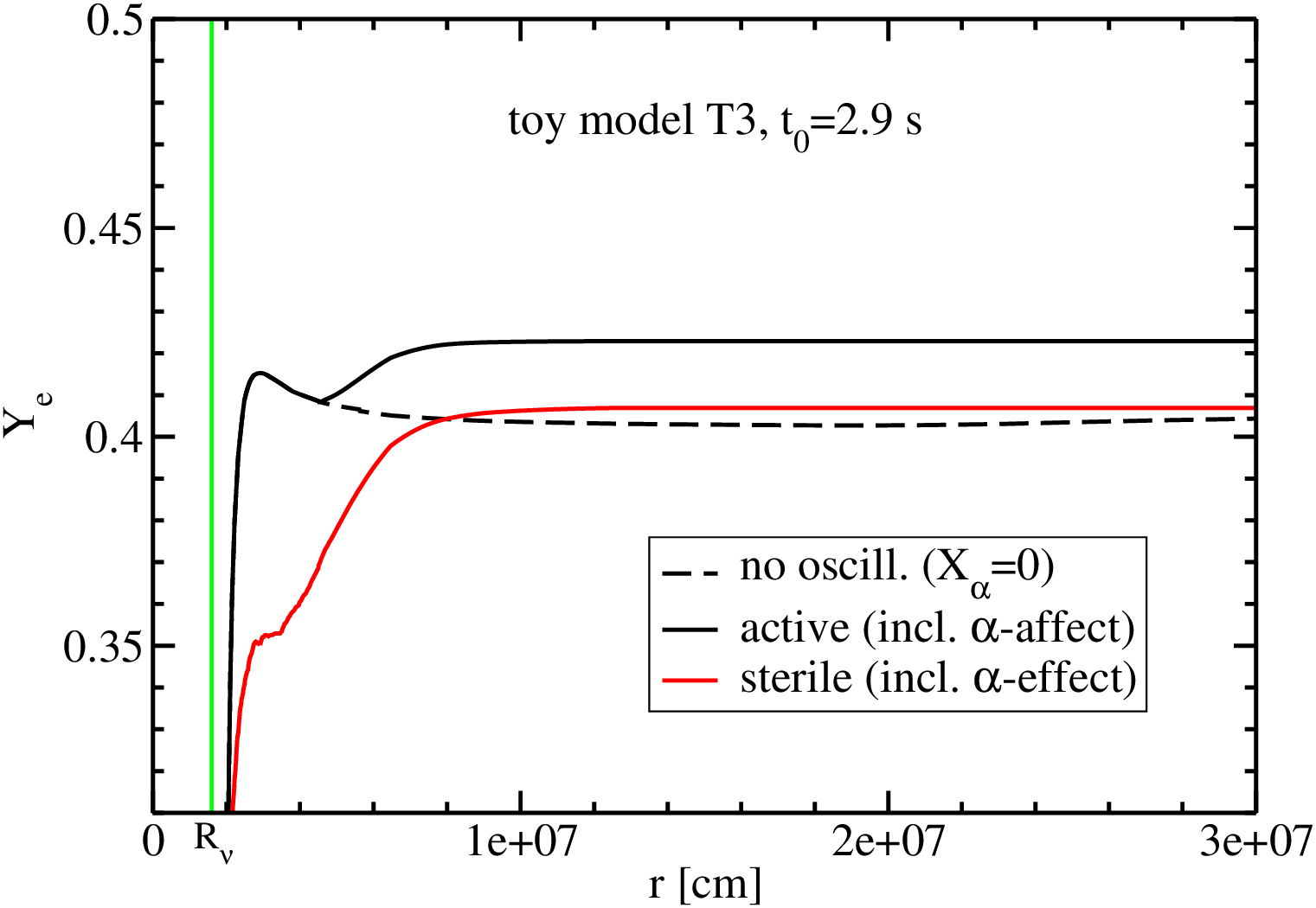}
\caption{Same as Fig.~\ref{fig:Toy1_Ye_2.9s}, but for toy model 3 (see text for details).
(The red lines here are running averages over $\Delta r \simeq 4.5\times 10^5$ cm.)}
\label{fig:Toy3_Ye_2.9s}
\end{figure*}
%--------------------------------------------------------------------------------

%--------------------------------------------
\begin{figure*}
%\plottwo{Toy3_6_5s_r_L_E.eps}{Toy3_r_Ye_6_5s.eps}
\plottwo{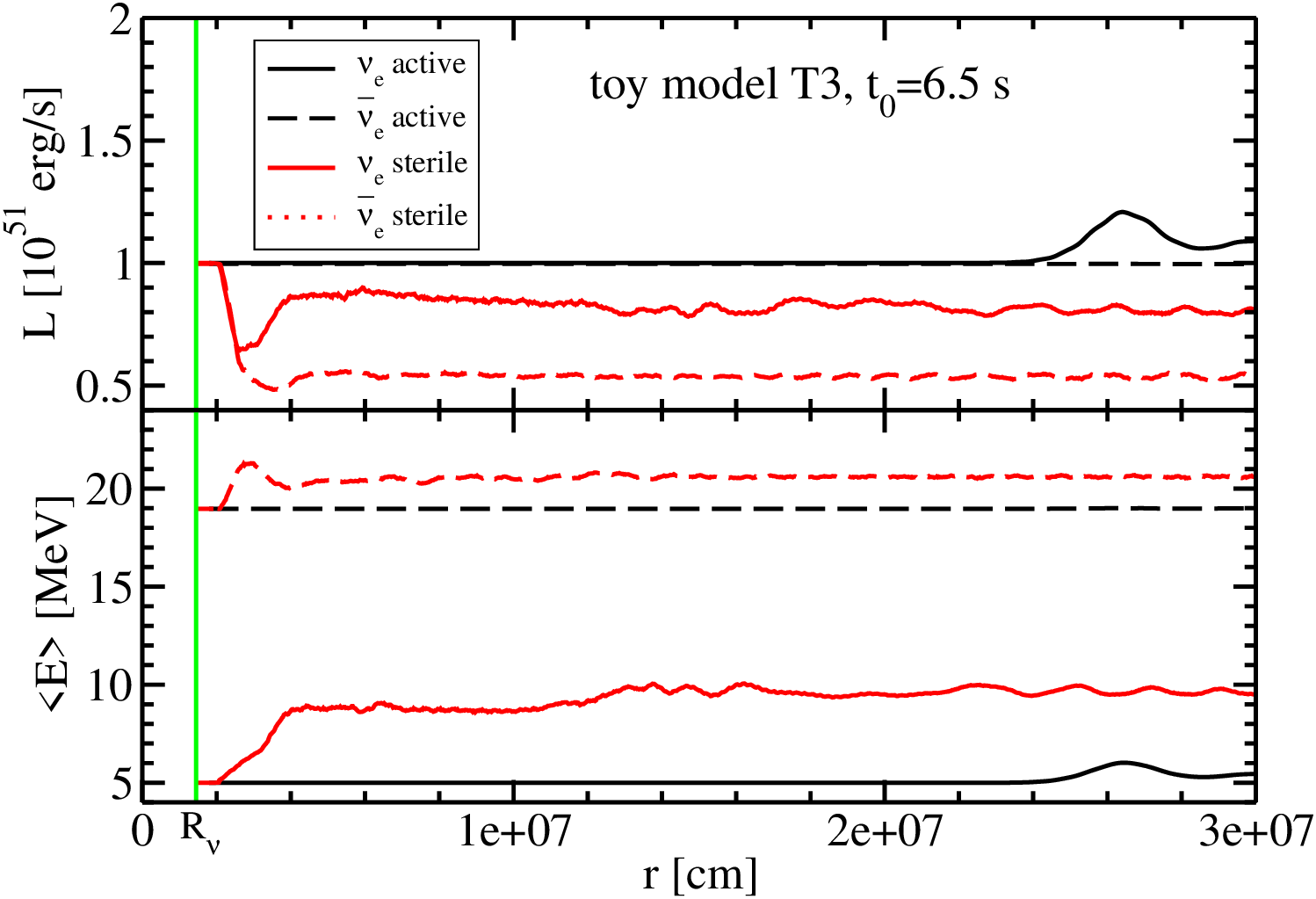}{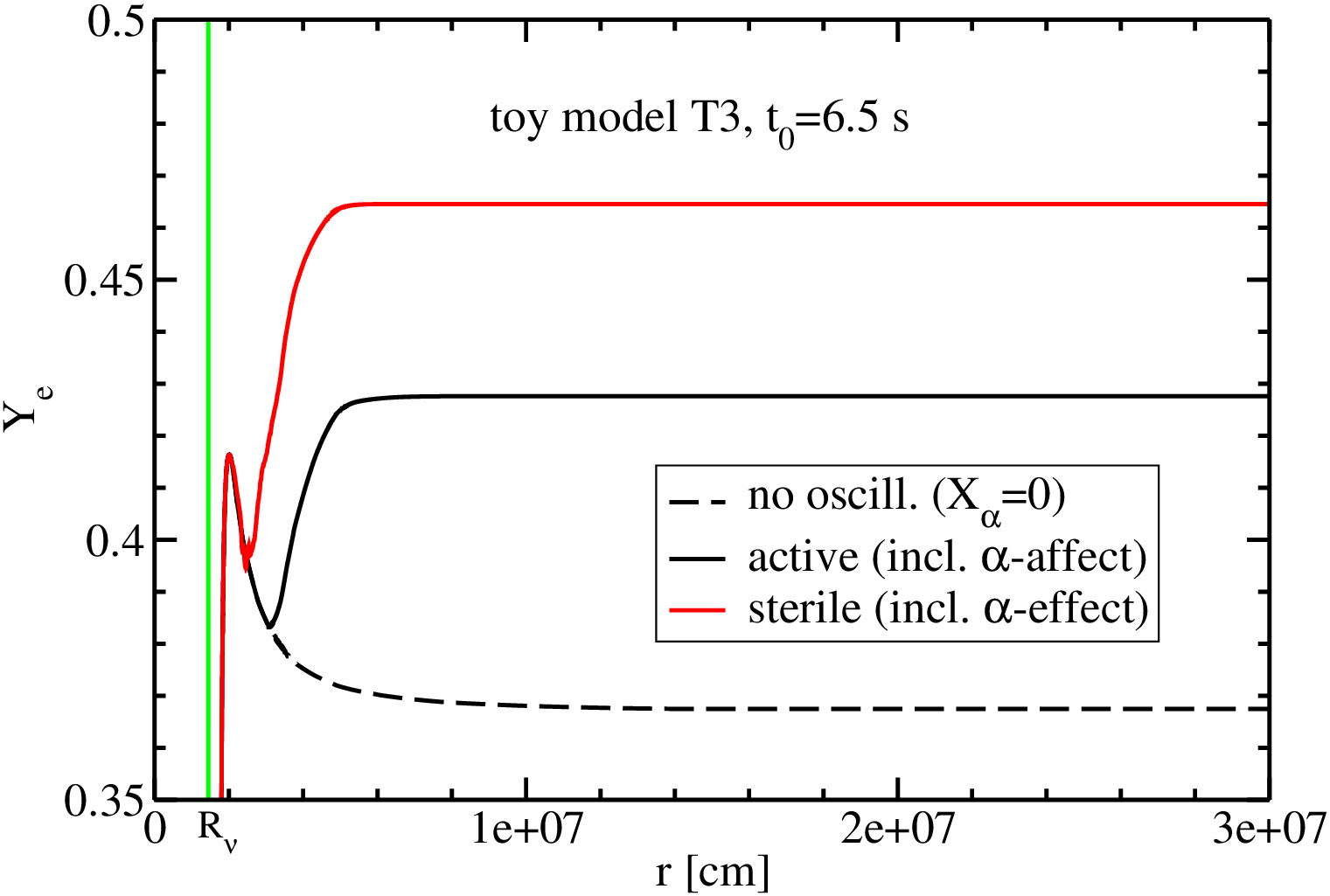}
\caption{Same as Fig.~\ref{fig:Toy1_Ye_6.5s}, but for toy model 3 (see text for details).
(The red lines here are running averages over $\Delta r \simeq 4.5\times 10^5$ cm.)}
\label{fig:Toy3_Ye_6.5s}
\end{figure*}
%--------------------------------------------------------------------------------

%-------------------------------------------------------------------------------
\begin{deluxetable*}{ccccccccccccccc}%{@{}c@{}l@{}c@{}c@{}c@{}c@{}c@{}c@{}c@{}c@{}c@{}c@{}c@{}l@{}l}
\tablecolumns{15}
\tabletypesize{\scriptsize}
%\tabletypesize{1pt}
%\setlength\columnsep{4pt}
\tablecaption{
Toy model parameters emulating mean-field nucleon 
potential corrections on the neutrino opacities\tablenotemark{a}.
\label{table:models_res}
}
\tablewidth{0pt}
\tablehead{\colhead{Toy}     &
 \colhead{$t_{0}$\tablenotemark{b}}                  &
 \colhead{$ L_{\nu_e} $  \tablenotemark{c}}             &
 \colhead{$ L_{\bar{\nu}_e} $ \tablenotemark{d}}                     &    
 \colhead{$ L_{\nu_x} $  \tablenotemark{e}}              &
  \colhead{$ L_{\nu_e}/\langle E_{\nu_e} \rangle$  \tablenotemark{f}}                 &
 \colhead{$ L_{\bar{\nu}_e}/\langle E_{\bar{\nu}_e} \rangle$ \tablenotemark{g}}         &
    \colhead{$ L_{\nu_x}/\langle E_{\nu_x} \rangle$  \tablenotemark{h}}                &   
  \colhead{$\langle E_{\nu_e} \rangle$  \tablenotemark{i}}                &
 \colhead{$\langle E_{\bar{\nu}_e} \rangle$ \tablenotemark{j}}                    & 
  \colhead{$\langle E_{\nu_x} \rangle$  \tablenotemark{k}}              &  
\colhead{$Y_{e,\mathrm{a}}$\tablenotemark{l}}                   &     
 \colhead{$Y^{X_{\alpha}=0}_{e,\mathrm{a}}$ \tablenotemark{m}}                         & 
\colhead{$Y^{\mathrm{act}}_{e,\mathrm{a}}$ \tablenotemark{n}}                       & 
\colhead{$Y^{\mathrm{ste}}_{e,\mathrm{a}}$ \tablenotemark{o}}                         \\ 
  \colhead{mod.}                                          &
  \colhead{[s]}                                                  &
  \colhead{[B/s]\tablenotemark{p}}                                                   &
  \colhead{[B/s]}                                                       &
  \colhead{[B/s]}                                                      &
  \colhead{[B/s]}                                                   &
  \colhead{[B/s]}                                                      &
  \colhead{[B/s]}                                                       &
  \colhead{[MeV]}                                                    &
  \colhead{[MeV]}                                                     &  
  \colhead{[MeV]}                                                    &
  \colhead{}                                                            &
    \colhead{}                                                     &  
  \colhead{}                                                       &
  \colhead{}                                                             
   }
\startdata
T1 & 2.9 & \textbf{3.30} & \textbf{3.40}  & \textbf{3.70} & 3.268 & 1.099 & \textbf{1.471} & 6.3 & 19.3 &  \textbf{15.7} & 0.422 &  0.403 &  0.422  & 0.430\\
T1 & 6.5 & \textbf{1.00} & \textbf{0.99}  & \textbf{1.04} & 1.248 &  0.325 & \textbf{0.549} & 5.0 & 19.0 & \textbf{11.8} &  0.428 &  0.368 &  0.428  &  0.510 \\
\hline
\hline \\[-0.07in]
T2 & 2.9 & 1.670 & 2.899 & \textbf{3.70} & \textbf{1.303}  & \textbf{1.302} & \textbf{1.471} & 8.0 & 13.9 & \textbf{15.7} & 0.420  & 0.405 & 0.421 & 0.440\\
T2 & 6.5 & 0.645 & 1.165 & \textbf{1.04} & \textbf{0.499}  & \textbf{0.518} & \textbf{0.549} & 8.0 & 14.0 & \textbf{11.8} & 0.431  & 0.380 & 0.431 & 0.486\\
\hline
\hline \\[-0.07in]
T3 & 2.9 & \textbf{3.30} & \textbf{3.40}  & \textbf{3.70}  & 3.268 & 1.099 & 1.196 & 6.3 & 19.3 & 19.3 & 0.422 & 0.403 & 0.422   &  0.407  \\
T3 & 6.5 & \textbf{1.00} & \textbf{0.99}  & \textbf{1.04}  & 1.248 & 0.325 & 0.342 & 5.0 & 19.0 & 19.0 & 0.428 & 0.368 & 0.428   &  0.465
\enddata
\tablenotetext{a}{In the first two cases (T1), we keep the neutrinospheric luminosities of $\nu_e$ 
and $\bar{\nu}_e$ as given by the hydrodynamical simulation, and  do not change the luminosity and 
mean energy of $\nu_x$ (see Table 1).
In the third and fourth cases (T2), we keep the neutrinospheric number fluxes of $\nu_e$ 
and $\bar{\nu}_e$ as given by the hydrodynamical simulation, and do not change the corresponding values
of $\nu_x$.
In the last two cases (T3), we keep the neutrinospheric luminosities of $\nu_e$, $\bar{\nu}_e$ 
and $\nu_x$ as given by the hydrodynamical simulation, and assume the same neutrinospheric mean energies
for $\bar{\nu}_e$ and $\nu_x$. 
Notice that in all cases we mark in boldface the unchanged hydrodynamical neutrinospheric parameters of $\nu_e$, $\bar{\nu}_e$ 
and $\nu_x$.}
\tablenotetext{b}{Postbounce time.}
\tablenotetext{c,d,e}{Neutrinospheric luminosities of $\nu_e$, $\bar{\nu}_e$ and $\nu_x$, respectively.}
\tablenotetext{f,g,h}{Neutrinospheric number fluxes of $\nu_e$, $\bar{\nu}_e$ and $\nu_x$, respectively.}
\tablenotetext{i,j,k}{Neutrinospheric mean energies of $\nu_e$, $\bar{\nu}_e$ and $\nu_x$, respectively.}
\tablenotetext{l}{Asymptotic wind electron fraction taking into account 
the $\alpha$-effect.}
\tablenotetext{m}{Asymptotic wind electron fraction without taking into account  
the $\alpha$-effect ($X_{\alpha}=0$).}
\tablenotetext{n}{Asymptotic wind electron fraction taking into account neutrino oscillations 
in the active sector and the $\alpha$-effect.}
\tablenotetext{o}{Asymptotic wind electron fraction taking into account neutrino oscillations
in the active and sterile sectors as well as the $\alpha$-effect.}
\tablenotetext{p}{1 Bethe = 1 B = 10$^{51}$ erg.}
\end{deluxetable*}
%-------------------------------------------------------------------------------

%%%%%%%%%%%%%%%%%%%%%%%%%%%%%%%%%%%%%%%%%%%%%%%%%%%%%%%%%%%%%%%%%%%%%%%%%%%%%%%%%
\section{Neutrino oscillations in a neutron-rich wind} \label{sec:toy-model}
%%%%%%%%%%%%%%%%%%%%%%%%%%%%%%%%%%%%%%%%%%%%%%%%%%%%%%%%%%%%%%%%%%%%%%%%%%%%%%%%%

In the previous sections, we considered the neutrino emission properties in the proton-rich 
environment obtained in the ECSN model of
\cite{Huedepohletal10}. As mentioned in Sect.~1, however, recent 
work suggests that these conditions might be valid only in the early ($t_\mathrm{pb} \lesssim 1$~s) and late
($t_\mathrm{pb} \gtrsim 3$~s) wind phases. Including mean-field nucleon 
potential corrections for charged-current neutrino opacities
in the dense medium of the proto-neutron star
\citep{Reddy1998} can cause $Y_{e}$ of the wind material 
to become neutron-rich \citep[possibly down to $Y_{e} \simeq
0.42$--0.45, see e.g.~][]{Roberts2012a, Martinez2012, Robertsetal12} during
an intermediate evolution period, although the
result is sensitively dependent on the employed nuclear equation of state. 
To explore the role of neutrino oscillations in such a
neutron-rich environment, we construct three toy models 
to emulate  mean-field corrections of the neutrino
opacities in their effect on lowering $\langle E_{\nu_e} \rangle$
and increasing $\langle E_{{\overline{\nu}}_e} \rangle$. Each
toy model case will be discussed  for an intermediate postbounce time ($t_{0} = 2.9$~s) and a late one ($t_{0} = 6.5$~s).

\subsection{Toy model inputs}
In all  toy models, we artificially prescribe the $\nu_e$ and $\bar{\nu}_e$ spectra by 
fixing the shape factors\footnote{We assume the shape factors of a moderately degenerate 
Fermi-Dirac distribution, for which  $\langle E^2_{\nu} \rangle/{\langle E_{\nu} \rangle}^2 \simeq 1.2$ 
\citep{HorowitzLi99}.}: 
$\alpha_{\nu_e} = \alpha_{\bar{\nu}_e} = 4$; 
The neutrino spectral properties not mentioned in the following
are assumed as in Table~1. 
 
In the first toy model (T1), we choose $\langle E_{\nu_e} \rangle$
and $\langle E_{{\overline{\nu}}_e} \rangle$ in order to obtain an 
asymptotic electron fraction \footnote{Note that \cite{Robertsetal12}
employed the approximate formula $Y_e\simeq1/(1+{\lambda_{\bar{\nu}_e}}/{\lambda_{\nu_e}})$ 
of \cite{QianFuller95} for estimating the electron fraction in the wind. 
This formula does not account for the $\alpha$-effect on $Y_e$.} 
including the $\alpha$-effect ($Y_{e,\mathrm{a}}$), or
neglecting it ($Y^{X_\alpha=0}_{e,\mathrm{a}}$), lower than 0.5  (see T1 in Table~\ref{table:models_res}). 
We then adopt the neutrino energy spectra and the electron fraction constructed in this way as initial
conditions to study the neutrino flavor evolution and its impact on the wind $Y_e$. 
The $\nu_x$ and $\bar{\nu}_x$ spectral properties are
unchanged (see Table 1).

Luminosities and mean energies simultaneously affect  $Y_e$. In order to prove the robustness of the T1 results, we consider another test case (toy model 2, T2), 
keeping the neutrinospheric number fluxes of $\nu_e$ and $\bar{\nu}_e$ (i.e., the $L_\nu/\langle E_\nu \rangle$ ratios) fixed as from the 
hydrodynamic simulation~\citep{Huedepohletal10} 
and varying both the luminosities and mean energies of $\nu_e$ and $\bar{\nu}_e$ in order to reproduce a neutron-rich environment 
in the absence of oscillations.
The new initial conditions are reported in Table~\ref{table:models_res} (case T2).
The third toy model (T3) is similar to T1, except that we assume 
$\langle E_{\nu_x}\rangle = \langle E_{\bar{\nu}_e}\rangle$ 
while leaving $L_{\nu_x}$ and $\alpha_{\nu_x}$ as in Table 1, in order to recover the usual hierarchy among the 
different neutrino flavors. 

\subsection{Neutrino oscillations}
Figures~\ref{fig:Toy1_Ye_2.9s} and \ref{fig:Toy1_Ye_6.5s} (left panels) 
show the luminosities and mean energies of $\nu_e$ and $\bar{\nu}_e$
in the active and sterile cases as functions of the radius for toy model 1.

In the active case,  the  initial conditions 
for neutrinos are different from the ones discussed in Sect.~\ref{sec:Ye_oscil} 
at the same $t_0$ (i.e., here we have $L_{\nu_e}/\langle E_{\nu_e}\rangle - L_{\nu_x}/\langle E_{\nu_x}\rangle > 0$). 
Moreover, the new spectral parameters also prescribe  larger differences between the $\nu_e$ ($\bar{\nu}_e$) and $\nu_x$
spectra and spectral crossings which are different from the previous cases. 
Bipolar oscillations due to $\nu$-$\nu$ 
interactions~\citep{Fogli:2008pt,Fogli2009JCAP} are then triggered at
$r \simeq 2.2 \times 10^7$~cm at $t_{0} = 2.9$~s and
$r \simeq 1.35 \times 10^7$~cm at $t_{0} = 6.5$~s. 
The neutrino and antineutrino luminosities and mean energies are correspondingly modified,  
as shown in Figs.~\ref{fig:Toy1_Ye_2.9s} and \ref{fig:Toy1_Ye_6.5s} (left panels). 

In the sterile case, at $t_0 =2.9$~s (see left panel of Fig.~\ref{fig:Toy1_Ye_2.9s}),
the inner active-sterile MSW resonance converts both $\nu_e$ and $\bar{\nu}_e$ to sterile states. 
As expected, $\nu_e$ are converted 
slightly more abundantly to sterile states than $\bar{\nu}_e$.   
Soon after, the ratio $L_{\nu_e}/\langle E_{\nu_e}\rangle$ increases, and the outer active-sterile MSW resonance 
occurs together with neutrino self-interactions. Note that due to the feedback effect on $Y_e$ and due to the 
initially lower value of $Y_e$ compared to 
the corresponding standard case, the outer MSW resonance is more adiabatic and it is expected to occur at smaller radii 
($r_{\rm OR} \simeq 4\times 10^6$~cm) than in the standard case. Moreover, due to the hierarchy  of the 
active neutrino fluxes and due to the lower matter potential, neutrino self-interactions  
mix $\nu_e$ and $\bar{\nu}_e$ with the heavy lepton flavors, increasing the $\nu_e$ survival probability, 
differently from what is shown in Fig.~\ref{fig:29s_L_E}. 

In the sterile case, at $t_0 =6.5$~s (see left panel of Fig.~\ref{fig:Toy1_Ye_6.5s}), 
the inner MSW resonance is visible as a  small drop of 
$L_{\nu_e}/\langle E_{\nu_e}\rangle$ (and even smaller for 
the $\bar{\nu}_e$) at $r_{\rm IR} \simeq 2\times 10^6$~cm. 
Slightly farther outside, at $r_{\rm OR} \simeq 2.5\times 10^6$~cm, 
the outer MSW resonance occurs (similarly to the standard case). Sterile neutrinos 
and antineutrinos are both abundantly produced through flavor conversions due to 
an interplay between the outer MSW resonance and collective oscillations, 
before $\alpha$ particles start forming at $r \simeq 3\times 10^6$~cm. 
As a consequence, both ${\nu}_e$ and $\bar{\nu}_e$ fluxes decrease,
causing an increase of $Y_e$ above 0.5 before the onset of the 
$\alpha$-effect. 
Toy model 2 is very similar to toy model 1 concerning the oscillation phenomenology, therefore we do not show our results 
here and only report the corresponding neutrino emission properties and 
the asymptotic $Y_e$ values in Table~\ref{table:models_res}.

Toy model 3 is shown in Figs.~\ref{fig:Toy3_Ye_2.9s} and \ref{fig:Toy3_Ye_6.5s}. 
In this case, the active flavors show a hierarchy of the mean energies more similar to the one reported in Table 1, 
although we have $L_{\nu_e}/\langle E_{\nu_e}\rangle - L_{\nu_x}/\langle E_{\nu_x}\rangle > 0$, 
similar to toy model T1. 
Therefore, in the active case, 
bipolar oscillations occur at $r > 3 \times 10^7$~cm at $t_{0} = 2.9$~s , while they start 
at $r \simeq 2.4 \times 10^7$~cm at $t_{0} = 6.5$~s.  
In the sterile case, the inner resonance is visible at $r_{\rm IR} \simeq 2 \times 10^6$~cm at $t_{0} = 2.9$~s. 
Soon afterwards $\nu$-$\nu$ interactions are triggered and the  $\nu_e$ survival probability starts  
to increase already before the region where the outer MSW resonance is expected 
to take place ($r_{\rm OR} \simeq 4 \times 10^6$~cm). 
In the sterile case, at $t_0 = 6.5$~s, the regions of the inner and outer MSW resonances 
almost overlap with each other, similarly to the standard case (Fig.~\ref{fig:6_5s_r_L_E}).
This is responsible for an overall drop of $L_{\nu_e}/\langle E_{\nu_e} \rangle$ and 
$L_{\bar{\nu}_e}/\langle E_{\bar{\nu}_e} \rangle$.

\newpage
\subsection{Feedback on the electron fraction}

In this section, we discuss the evolution of the electron fraction for
toy models T1, T2, and T3 considered in the previous section in order to disentangle 
between the impact of neutrino oscillations and $\alpha$-effect 
on $Y_e$ in a neutron-rich neutrino-driven wind environment.

In Figs. \ref{fig:Toy1_Ye_2.9s} and \ref{fig:Toy1_Ye_6.5s} (right panels),
we show the evolution of $Y_{e}$ at intermediate ($t_{0}=2.9$ s) and 
late ($t_{0}=6.5$ s) evolution phases of the neutrino-driven wind in the T1 
model (see Table 2). The dashed lines refer to $Y_{e}$ in the case where neither neutrino
oscillations nor the $\alpha$-effect are taken into account,
while the solid lines display $Y_e$ radial evolutions, including the $\alpha$-effect 
in the active (solid black lines) and sterile  (solid red lines) cases.
Since active oscillations take place at $r > 1.2 \times 10^7$~cm in 
both cases (i.e., after $Y_e$ has reached its asymptotic value),  
the difference between ${Y}^{\mathrm{act}}_{e,\mathrm{a}}$ and $Y^{X_{\alpha}=0}_{e,\mathrm{a}}$ 
is just caused by the $\alpha$-effect (see Table 2), which pushes $Y_e$
towards 0.5, as expected.  
The impact of the $\alpha$-effect on $Y_e$ is larger 
at late times also in these toy models, 
for the reasons we already discussed in Sect.~6.
 
In the sterile case, neutrino oscillations raise the asymptotic value 
of the electron fraction compared to the active case, therefore  
the matter becomes \textit{more proton-rich} compared to the case where
oscillations are not considered or where they occur in the active sector
only. 

In particular, at $t_{0}=2.9$ s, the inner and outer MSW resonances in the sterile
case cause $Y_e$ to be lower than in the active case, already before $\alpha$ particles 
start forming. 
Then, the $\nu$-$\nu$ interactions, which repopulate the $\nu_e$ sector, 
drive $Y_e$ towards 0.5 and even above the value of 
$Y_e$ in the active case, even without the $\alpha$-effect,
which removes free nucleons and thus moderates
the impact of neutrino oscillations on $Y_e$,
as discussed in detail in Sect.~6. 

At $t_{0}=6.5$ s, neutrino oscillations occur very close to the neutrinosphere and
push $Y_e$ in the sterile case to a much higher value ($> 0.5$) than in the active case,
already before $\alpha$ particles start forming at $r\sim 4.0\times 10^{6}$~cm. 
Therefore, the formation of $\alpha$ particles impacts the evolution of $Y_e$ 
in the usual way, namely towards more symmetric conditions ($Y_e\longrightarrow0.5$).

In order to prove the robustness of our conclusions 
about the interplay between neutrino oscillations and the $\alpha$-effect,
we also calculate $Y_{e}$ for toy models T2 and T3.
We do not show the evolution of $Y_e$ for T2, because the discussion
is very similar to T1, but we report the corresponding $Y_e$ results in Table 2.

In the T3 active case (see black solid lines in Figs.~\ref{fig:Toy3_Ye_2.9s} and 
\ref{fig:Toy3_Ye_6.5s}), the discussion about the impact of the $\alpha$-effect 
and neutrino oscillations on the evolution of $Y_e$ at $t_0=2.9$ s 
and $t_0=6.5$ s is very similar to what we already discussed in the T1 case. 

In the T3 sterile case, instead, at $t_{0}=2.9$ s, we observe an interesting 
interplay between neutrino oscillations and the $\alpha$-effect, 
because $Y_e$ in the sterile case (solid red line) 
is lower than in the active case (solid black line), 
different from cases T1 and T2 at $t_{0}=2.9$ s. This is due to the fact 
that the MSW resonances initially deplete the number flux of $\nu_e$
in favor of $\nu_s$ much more than in cases T1 and T2,
for the reasons discussed in the previous subsection. 
Therefore, $Y_{e}$ in the sterile case is already
much lower than in the active case,
before $\alpha$ particles start forming. 
In the following evolution, different from cases T1 and T2,
the $\alpha$-effect damps the efficiency 
of $\nu$-$\nu$ interactions in raising $Y_{e}$. 
Therefore, the latter remains lower than in the active case. 
However, the difference between $Y_{e}$ in the active case and 
in the sterile case is not sufficiently large  
to conclude that neutrino oscillations in the 
sterile case make the environment significantly
more neutron-rich than in the case without neutrino oscillations.

At $t_{0}=6.5$ s, we basically observe the same trend as 
in cases T1 and T2, namely $Y_{e}$ in the sterile
case is higher than in the active case. The reasons 
are very similar to what was already discussed for the 
T1 model.

In conclusion, neutrino oscillations (with or without sterile
neutrinos) combined with the $\alpha$-effect do not support very neutron-rich
conditions in the neutrino-driven wind for the considered SN model. 
Conditions for a strong r-process in this SN progenitor are disfavored, because 
$Y_{e}$ tends to be pushed close to 0.5 and thus the formation of a
highly neutron-rich environment is prevented.

%%%%%%%%%%%%%%%%%%%%%%%%%%%%%%%%%%%%%%%%%%%%%%%%%%%%%%%%%%%%%%%%%%%%%%%%%%%%%%%%%
\section{Discussion} \label{sec:discussion}
%%%%%%%%%%%%%%%%%%%%%%%%%%%%%%%%%%%%%%%%%%%%%%%%%%%%%%%%%%%%%%%%%%%%%%%%%%%%%%%%%
In this work, we studied the nucleosynthesis outcome of an ECSN with 8.8\,$M_{\odot}$, by adopting the SN model presented in~\cite{Huedepohletal10}. 
The same SN simulation was adopted in \cite{Tamborraetal12} to study the impact of neutrino oscillations on the electron fraction 
in the presence of light sterile states. However, due to the complications
induced by the numerical solution of a large number of non-linear, coupled equations
with three neutrino families and the oscillation feedback on $Y_e$,  the inner MSW resonance
was not included in~\cite{Tamborraetal12}, assuming that its impact on the electron fraction was negligible
during the neutrino-driven wind phase due to the steepness of the matter potential in that region. 
It was found that neutrino conversions to a sterile
flavor and neutrino self-interactions influence the radial asymptotic
value of $Y_e$ in the neutrino-driven wind in complicated and time-dependent ways.
These conclusions
motivated us to investigate in detail the effect of oscillations on 
a larger variety of wind conditions and on
the nucleosynthetic abundances.  

In this work, the neutrino evolution is followed from the neutrinosphere outward. 
We also develop a more detailed treatment of the $Y_e$ evolution than in 
\cite{Tamborraetal12}, by accounting for the $\alpha$-effect as well as  recoil and weak magnetism 
corrections  in the $\beta$ processes. 
We find that the inner active-sterile MSW resonance has a negligible impact on $Y_e$ during the intermediate and late 
cooling phases, although  it 
modifies the $\nu$ and $\bar{\nu}$ spectra. In particular, as discussed in Appendix~\ref{sec:AppendixA}, when $\nu$-$\nu$ interactions 
are included,  the flavor instability 
induced by the active-sterile MSW resonance triggers neutrino self-interactions modifying the 
flavor evolution history compared to the case where only interactions  with the matter background are considered.
On the other hand, the inner MSW resonance induces non-negligible modifications of the electron fraction 
during the accretion phase, as
pointed out in~\cite{Wuetal2013}, and in the early proto-NS cooling phase. It is responsible for 
the formation of a plateau in $Y_e$ that drives the asymptotic value of $Y_e$ towards smaller values. 

The early cooling phase (i.e., at $t_0 = 0.5$~s and $1$~s) was also 
discussed in \cite{Wuetal2013} for the same ECSN progenitor, but adopting the simulation of ~\cite{Fischeretal09}. 
Including sterile neutrino oscillations \cite{Wuetal2013} obtain a neutron-rich environment ($Y_e^a [0.5 {\rm s}] = 0.38$)  
differently from our results ($Y_e^a [0.5 {\rm s}] = 0.53$, see Figs.~\ref{fig:05s_L_E} and \ref{fig:TOT_r_Ye}). 
Such a discrepancy might be due to the different supernova models adopted as inputs  in \cite{Wuetal2013} (i.e., \citealp{Fischeretal09}) and in our work 
(i.e., \citealp{Huedepohletal10}). In fact,  
the electron fraction without oscillations is $Y_e [0.5 s]=0.49$ in Fig.~3 (red curve) of \cite{Wuetal2013}, while in our case it is $Y_e [0.5s]=0.56$ as 
shown in Fig.~\ref{fig:05s_L_E}.  
Our work also adopts an approach to study the electron
fraction evolution different from the one employed in \cite{Wuetal2013} 
(i.e., our Eq.~\ref{Yeeq} vs. Eq.~5 of \citealp{Wuetal2013}). The static approach 
of~\cite{Wuetal2013}  carries ``memory'' of the large modifications of the neutrino
fluxes and of the electron fraction due to the inner MSW resonance at $t < 0.5$~s, 
while our sampling is sparse, because  $90\%$ of the ejecta of the 
early cooling are combined into one trajectory  ejected at $0.5$~s. 
Even adopting a denser grid in $t_0$, our dynamic approach should not be accurate 
during the accretion phase where the steady-state approximation is not applicable. 
Other differences on $Y_e$ might be due to a different treatment of the neutrino oscillations. In \cite{Wuetal2013}, a $1({\rm active}) + 1({\rm sterile})$ 
approximation is adopted 
and $\nu$-$\nu$ interactions are neglected assuming that they are suppressed due to the high
matter potential during the accretion phase~\citep{Sarikas2012PhRvL}, while we include the $\nu_e$-$\nu_x$ flavor mixing as well as 
neutrino self-interactions in our computations.

Given the complex and nonlinear nature of neutrino self-interactions, all existing numerical studies with 
neutrino-neutrino refraction use simplifying  assumptions.
In our treatment of the neutrino evolution, we averaged the angular dependence of $\nu$-$\nu$
interactions (the so-called ``single-angle approximation,''~\citealp{Duan:2006an}). 
Because of the similarity between the $\nu_e$ and $\bar{\nu}_e$ fluxes
and those of the corresponding heavy-lepton neutrinos in our 
hydrodynamical simulations, and because of the observed strength of the $\alpha$-effect 
in pushing $Y_e$ close to 0.5, even a possible relevance of 
multi-angle effects due to a small asymmetry among the neutrino fluxes of different flavors~\citep{EstebanPretel:2007ec} 
is unlikely to play any important role for $Y_e$. In the sterile case, the asymmetry between $\nu_e$ and $\nu_x$
becomes even larger than in the active case due to the $\nu_s$ production, therefore we expect that a 
full-multi-angle treatment would only induce a smearing
of the neutrino fluxes~\citep{Fogli:2007bk}, without a dramatic impact on $Y_e$. 
If the matter potential is high enough, neutrino multi-angle effects could also be 
responsible for a matter suppression of collective effects, and therefore
produce results which are different from the ones obtained within the ``single-angle'' approximation~\citep{EstebanPretel:2008ni}. 
A multi-angle study was developed by~\cite{Chakraborty:2011gd} for  one energy mode and for the $8.8\ M_\odot$ progenitor 
presented in  \cite{Fischeretal09}: A  complete matter suppression 
of the collective effects due to multi-angle matter effects was never achieved for this progenitor, 
because of the low-density matter profile.
We therefore suspect that also the triggering of the collective effects induced by the inner MSW resonance instability should not be suppressed
by a multi-angle treatment of the neutrino flavor oscillations during the cooling phase. 
Nevertheless, more accurate studies including multi-angle effects are 
mandatory and should be conducted for  a larger sample of supernova progenitors and nuclear equations of state, especially because, according to the modeling  
presented in \cite{Duanetal11},  it was concluded that multi-angle effects among active flavors may affect the nucleosynthetic outcome under certain conditions. 

Concerning the nucleosynthesis outcome, in the case without sterile neutrino and neutrino oscillations,
all relevant results can be found in Fig.~3 of \cite{Wanajo2011aaa}: There are nucleosynthetic yields for 
a 1D model (to be directly compared with  \citealp{Wuetal2013}) and of a more
realistic 2D model as well. The 2D model yields major and important differences compared to the 1D case,
as in all details discussed in \cite{Wanajo2011aaa}. The differences between 
the 1D nucleosynthesis result of \cite{Wanajo2011aaa} and \cite{Wuetal2013} are most probably due to differences
in the mass-versus-$Y_e$ distribution, which are probably caused by differences in the 
neutrino interaction processes adopted in our models compared to those of \cite{Fischeretal09}.
Since the mass-versus-$Y_e$ distribution is provided only by \cite{Wanajo2011aaa}
but not in \cite{Wuetal2013}, a detailed 
comparison between our and their nucleosynthesis results is not possible.

%%%%%%%%%%%%%%%%%%%%%%%%%%%%%%%%%%%%%%%%%%%%%%%%%%%%%%%%%%%%%%%%%%%%%%%%%%%%%%%%
\section{Conclusions}
%%%%%%%%%%%%%%%%%%%%%%%%%%%%%%%%%%%%%%%%%%%%%%%%%%%%%%%%%%%%%%%%%%%%%%%%%%%%%%%%%
We presented neutrino oscillations and nucleosynthesis calculations for the neutrino-cooling phase of 
the proto-neutron star born in an 8.8\,$M_{\odot}$ electron-capture supernova, 
using trajectories for the $\nu$-driven wind from 1D hydrodynamic 
simulations, in which a sophisticated treatment of neutrino transport
was applied \citep{Huedepohletal10}. 
In particular, we studied the consequences of neutrino oscillations 
of two active flavors 
driven by the atmospheric mass difference and $\theta_{13}$ and, 
motivated by  hints on the possible
existence of light sterile neutrinos, we also discussed the role of flavor
oscillations with 1 sterile + 2 active flavors.
In our study neutrino-neutrino refraction effects were included, too.
We chose $\nu_e$-$\nu_s$ mixing
parameters as suggested by the reactor anomaly~\citep{Mention:2011rk}. 
However, our conclusions remain valid also for moderate variations of
the sterile mass-mixing parameters. 

Our results demonstrate that the $\alpha$-effect plays a crucial 
role in discussing the consequences of neutrino oscillations on the
$Y_e$ evolution in neutrino-driven winds. It can damp as well as
enhance the $Y_e$-reducing impact of $\nu_e$-$\nu_s$ conversions,
depending on the radial position of the active-sterile MSW region
relative to the radius where $\alpha$ particles form from nucleon
recombination. In the late proto-neutron star cooling phase 
the production of sterile neutrinos via an MSW resonance takes place very close to the
neutrinosphere, while a significant abundance of $\alpha$ particles 
in the wind appears only at larger distances. The $Y_e$ reduction 
in the ejecta associated with the transformation of $\nu_e$ to 
$\nu_s$ is therefore amplified by the subsequent $\alpha$-effect,
driving $Y_e$ from initial values considerably above 0.5 to
an asymptotic value closer to 0.5. In the early wind phase the
effect is different. Here the outer $\nu_e$-$\nu_s$ MSW conversions
occur farther away from the neutron star and exterior to
(or coincident with) the formation region of $\alpha$ particles.
The $\alpha$-effect then {\rm moderates} the $Y_e$ reduction caused
by the presence of sterile neutrinos. Because of this dominance of 
the $\alpha$-effect, the asymptotic neutron-to-proton ratio in the 
early wind becomes very similar in the cases with and without 
sterile neutrinos (whereas without the $\alpha$-effect sterile neutrinos 
always cause a significant reduction of $Y_e$).

While the neutrino-driven wind of our ECSN model is well on the
proton-rich side \citep{Huedepohletal10}, equation-of-state dependent
nucleon mean-field potentials in the neutrinospheric region might
lead to a considerably lower $Y_e$ in the wind outflow 
\citep{Robertsetal12,MartinezPinedoetal11}. 
For this reason we constructed six toy model cases for the
intermediate and late wind phases, in which the (unoscillated) neutrino spectra were 
chosen such that the neutrino-driven wind became neutron-rich
with an asymptotic wind-$Y_e$ (including the $\alpha$-effect) 
 of about 0.42--0.43, which is on the extreme side of the theoretical estimates.
Including active-sterile flavor oscillations, the outflow turns, in some cases, 
{\em more proton-rich}, despite the conversion of $\nu_e$ to $\nu_s$.
This counterintuitive $Y_e$ increase is caused by neutrino oscillations,  
which modify the neutrino emission properties such that either the 
$\nu_e$ absorption is more strongly increased than the competing 
$\bar\nu_e$ absorption or the $\bar\nu_e$ absorption is more 
strongly reduced than the competing $\nu_e$ absorption.
Our conclusion that sterile neutrinos are unlikely to help
enforcing neutron-rich conditions in the wind ejecta therefore
seems to remain valid even when nucleon-potential effects are taken
into account in future neutron-star cooling simulations.

If oscillations are disregarded, the wind ejecta in our ECSN model
develop a proton excess and therefore
only iron-group and some $p$-rich isotopes are created with small
production factors (below 10), not adding any significant production
of interesting isotopes to the nucleosynthesis yields computed for
the early ejecta of 2D explosion models of such ECSNe
\citep{Wanajo2011aaa,Wanajo2013,Wanajoetal13b}. 
When neutrino oscillations are taken
into account by our simplified neutrino-mixing scheme, the
feedback of oscillations on $Y_e$ is time-dependent, since it is
sensitive to the detailed matter profile and neutrino fluxes. 
In the early $\nu$-driven wind, the asymptotic $Y_e$ value
in the presence of a sterile family is lower 
than the $Y_e$ value obtained without oscillations, although 
always $>0.5$. In the intermediate phase of the $\nu$-driven wind
$Y_e$ in the presence of sterile neutrinos is even a bit higher 
than the one without oscillations.
In the late $\nu$-driven wind
the asymptotic $Y_e$ in the presence of sterile neutrinos
is slightly lowered compared to the case without oscillations or
to the case where oscillations in the active sector are considered.
However, in our model of the neutrino cooling of the proto-neutron
star born in an ECSN, the corresponding effects do not lead to any
neutron excess. The changes of the nucleosynthetic output for
models with (active or sterile) neutrino oscillations compared to
the no oscillations case are 
insignificant. It appears unlikely that in the studied progenitor
viable conditions for strong r-processing can be established.

Our conclusions concern the $\nu$-driven wind of an 8.8\,$M_{\odot}$
progenitor. More studies of the impact of neutrino oscillations on the
early-time ejecta 
including multi-dimensional effects arising in hydrodynamic
simulations~\citep{Wanajo2011aaa,Wanajo2013} and including
the effects of nucleon mean-field potentials in the neutrino opacities,
are needed in order to shed light on the consequences of neutrino
oscillations for the explosion mechanism and nucleosynthetic abundances
(cf. \citealp{Wuetal2013}, who considered only a 1D model).
Studies of a broader range of progenitor
models, in particular also iron-core SNe with more massive proto-neutron
stars, applying state-of-the-art neutrino-oscillation
physics, are also desirable to identify possible cases where favorable
conditions for an r-process may be produced.

%%%%%%%%%%%%%%%%%%%%%%%%%%%%%%%%%%%%%%%%%%%%%%%%%%%%%%%%%%%%%%%%%%%%%%%%%%%%%%%%%
\acknowledgements 
\section*{Acknowledgments}
%%%%%%%%%%%%%%%%%%%%%%%%%%%%%%%%%%%%%%%%%%%%%%%%%%%%%%%%%%%%%%%%%%%%%%%%%%%%%%%%% 
%E.P. is thankful to Bernhard M\"uller for useful discussions.
{We thank Bernhard M\"uller and Meng-Ru Wu for useful discussions.}
I.T. acknowledges support from the Netherlands Organization for Scientific
Research (NWO). S.W. acknowledges partial support from the RIKEN iTHES Project
and the JSPS Grants-in-Aid for Scientific Research (26400232, 26400237).
Partial support from the Deutsche Forschungsgemeinschaft 
through the Transregional Collaborative Research Center SFB/TR 7
``Gravitational Wave Astronomy'' and the Cluster of Excellence EXC 153 
``Origin and Structure of the
Universe'' (http://www.universe-cluster.de)
is also acknowledged.

\newpage
\appendix
%%%%%%%%%%%%%%%%%%%%%%%%%%%%%%%%%%%%%%%%%%%%%%%%%%%%%%%%%%%%%%%%%%%%%%%%%%%%%%%%%%%%%%%%%%%%%%%%%%%%%%%%%%%%
\section{Feedback of neutrino self-interactions on the electron fraction} \label{sec:AppendixA}%
%%%%%%%%%%%%%%%%%%%%%%%%%%%%%%%%%%%%%%%%%%%%%%%%%%%%%%%%%%%%%%%%%%%%%%%%%%%%%%%%%%%%%%%%%%%%%%%%%%%%%%%%%%%%
Given the non-linear nature of neutrino self-interactions, in this appendix we 
discuss the oscillation physics at the selected postbounce times 
$t_0 = 2.9$ and $6.5$~s where $\nu$-$\nu$ interactions  
significantly affect the neutrino spectral properties.   
In order to disentangle between the role played by $\nu$-$\nu$ interactions 
and the MSW resonances, we also discuss a simpler case obtained by 
switching off the neutrino self-interaction term (i.e., including  matter effects only).  
Note that here, we discuss  the sterile case with the prescription (ii) in Sect.~3 
for $\alpha$ particles, namely without taking into account the $\alpha$-effect. 

A simple quantity that can be introduced in order to have an idea about the 
locations of the active-sterile MSW resonances is  the refractive  energy difference between $\nu_e$ 
and $\nu_s$ caused by matter and neutrino refraction (see Sect.~\ref{sec:neutrinos}):

%....................................................
\begin{eqnarray}\label{lambdaes}
V_{es} = {\sf H}^{{\rm m}+\nu\nu}_{ee} - {\sf H}^{{\rm m}+\nu\nu}_{ss}
&=&\sqrt{2}G_{\rm F}\left[N_b\left(\frac{3}{2} Y_e - \frac{1}{2}\right) + 2 (N_{\nu_e} - N_{\bar\nu_e}) + (N_{\nu_x} -N_{\bar\nu_x}) \right]\ .
\end{eqnarray}
%....................................................

\begin{figure*}
\resizebox{1.0\textwidth}{!}{
\includegraphics*{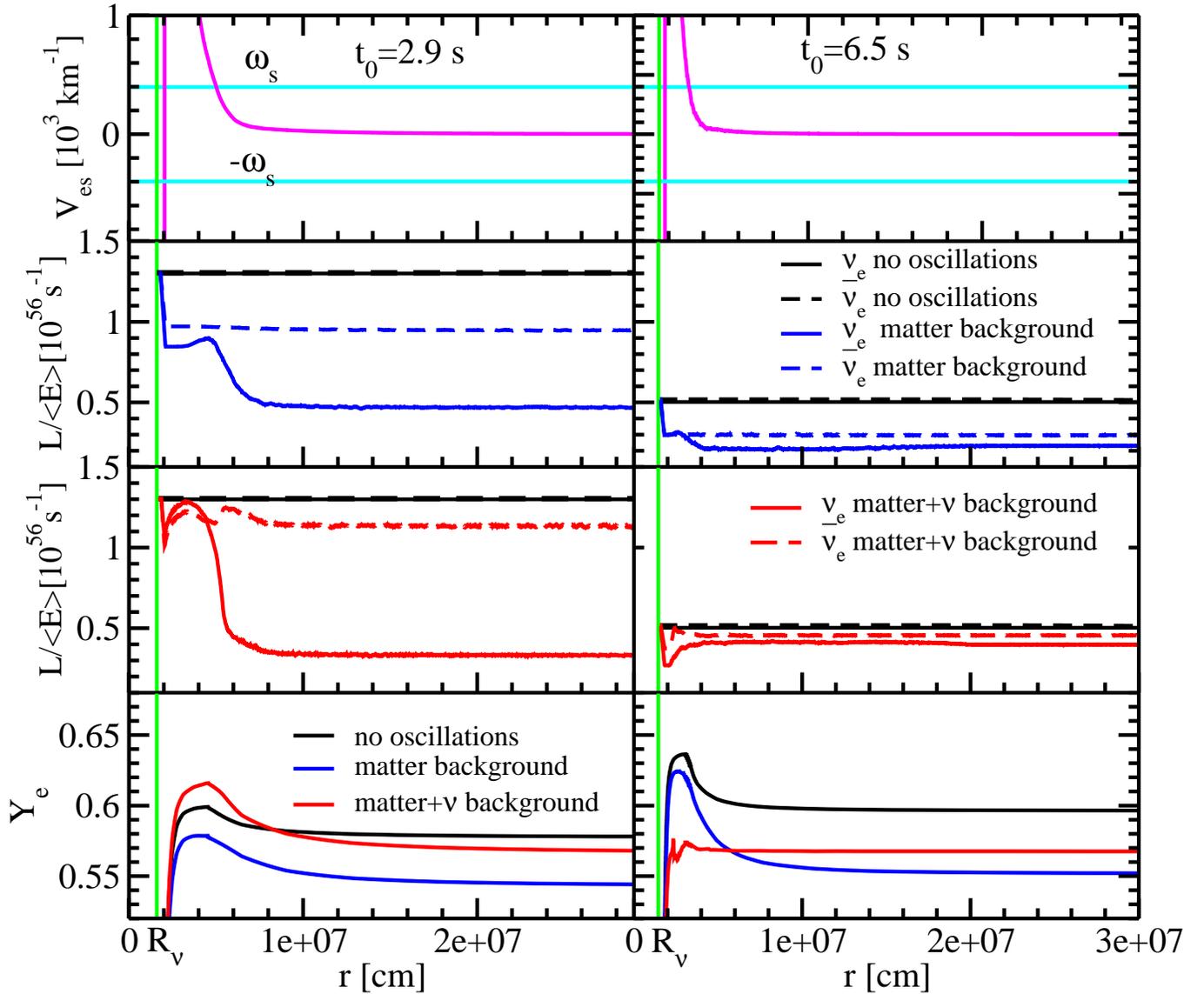}}
\caption{{\em Top 
 panel:} Refractive energy difference between $\nu_e$ and $\nu_s$ ($V_{es}$) in the sterile case. The horizontal 
 lines ($\pm \omega_{S}$) mark the oscillation frequency of a typical energy mode of $15$~MeV for neutrinos and antineutrinos. 
 {\em Second panel:} $L_{\nu_e}/\langle E_{\nu_e} \rangle$ and $L_{\bar{\nu}_e}/\langle E_{\bar{\nu}_e} \rangle$   
 as functions of the distance ($r$) from the center of
 the PNS in the sterile case with matter background and in the case without oscillations. {\em Third panel:} 
 $L_{\nu_e}/\langle E_{\nu_e} \rangle$ and $L_{\bar{\nu}_e}/\langle E_{\bar{\nu}_e} \rangle$ in the sterile case with 
 matter$+\nu$ background and in the case without oscillations. {\em Fourth panel:}
 Electron fraction $Y_{e}$ as a function of the
distance $r$ from the center of the PNS in the matter background, matter $+ \nu$ background and no oscillation cases, 
without including the $\alpha$-effect. The panels on the left side refer to $t_{0}=2.9$ s postbounce time, while the ones on the right to $t_{0}=6.5$ s.
The vertical line marks the neutrinosphere radius $R_\nu$.
}
\label{fig:Ves}
\end{figure*}

We show $V_{es}$ (Eq.~\ref{lambdaes}) as a function of the radius at
$t_0 = 2.9$~s in the top left panel of Fig.~\ref{fig:Ves}.  
This profile already includes a self-consistent solution of $Y_e$.   
The regions where we should expect the inner and outer active-sterile MSW resonances 
are defined by the intersection of the $V_{es}$ profile with the $\pm \omega_S$ lines 
(corresponding to the typical oscillation frequency of $15$~MeV $\nu$ and $\bar{\nu}$, see Eq.~22): 
The  MSW resonances should occur at $r_{\rm IR} \simeq 2 \times 10^6$~cm for $\nu$ and $\bar{\nu}$ 
and at $r_{\rm OR} \simeq 4.5\times10^{6}$~cm for neutrinos only.

In the matter background case,  the $\nu_e \rightarrow \nu_s$ conversions
are more abundant than the antineutrino ones, as already discussed in Sect.~\ref{sec:Ye_oscil}. 
Correspondingly, the electron fraction  
(fourth panel of Fig.~\ref{fig:Ves} on the left) is lower than the one in the case without oscillations. 
The outer resonance occurs at $r_{\rm OR} \simeq 50 \times 10^5$~cm 
only for neutrinos and it favors an even lower value of the electron fraction. 

In the matter$+\nu$ background case, the inner MSW resonance takes place  at 
the same radius as in the matter background case (see second and third panels on
the left of Fig.~\ref{fig:Ves}), but  
a slightly lower fraction of the $\nu_e$ are converted to $\nu_s$ because of 
${\sf H}^{\nu\nu} \neq 0$. Moreover, it is clear by 
comparing the second and the third panels of Fig.~\ref{fig:Ves}, that
the $\nu$-$\nu$ interaction term is responsible for 
replenishing the $\nu_e$ flux before the outer active-sterile MSW 
resonance occurs at $r_{\rm OR} \simeq 5 \times 10^6$~cm. 
Correspondingly, the electron fraction increases compared to the 
case without oscillations close to the neutrinosphere and decreases 
afterwards because of the outer MSW resonance. 
Comparing the $Y_e$ profiles in the matter$+\nu$ background and 
matter background cases, we find that $Y_e(\mathrm{matter}+\nu) - Y_e(\mathrm{matter})\simeq0.025$. 

The right side of Fig.~\ref{fig:Ves} shows the same quantities discussed at 
$t_0 = 2.9$~s, but at $t_0=6.5$~s. 
In the matter background case,  the inner MSW resonance occurs at $r_\mathrm{IR} \simeq 1.8 \times 10^6$~cm as shown by   
the $L_{\nu_e}/\langle E_{\nu_e} \rangle$ and $L_{\bar{\nu}_e}/\langle E_{\bar{\nu}_e} \rangle$
behavior in the second panel of Fig.~\ref{fig:Ves} and by the top panel of the same figure. In this case, almost the 
same amount of $\nu_e$ and $\bar{\nu}_e$ is converted
to sterile states (due to the steepness of the matter 
potential as discussed in Sect.~\ref{sec:Ye_oscil}). The outer MSW resonance occurs closer to the inner one than at  
$t_0= 2.9$~s (at $r_\mathrm{OR} \simeq 2.5 \times 10^6$~cm)
and it is responsible for depleting the $\nu_e$ flux in favor of sterile state production. Therefore, the electron fraction, 
plotted in the fourth right panel of Fig.~\ref{fig:Ves},
becomes lower than in the case without oscillations. 

In the matter$+\nu$ background case, the role played by the neutrino self-interactions
is evident already close to the inner resonance. In fact, the difference between the $\nu_e \rightarrow \nu_s$ 
and $\bar{\nu}_e \rightarrow \bar{\nu}_s$ flavor conversions is responsible for lowering $Y_e$ compared to the case without oscillations
and with matter background only. Soon afterwards, and in correspondence to the outer resonance, 
the interplay between the matter and neutrino background and the 
non-linear effects due to $\nu$-$\nu$ interactions is responsible for partially repopulating the $\nu_e$ and  the $\bar{\nu}_e$ sectors and, 
as a consequence, $Y_e$ does not decrease further as it happens in the case at $t_0 = 2.9$~s.  Comparing the $Y_e$ profiles in the matter$+\nu$ 
background and matter background cases, we find that $Y_e(\mathrm{matter}+\nu) - Y_e(\mathrm{matter})\simeq0.02$.  
For both the discussed profiles, $\nu$-$\nu$ interactions are triggered at smaller radii than usually expected in the active case 
by the presence of non-zero off-diagonal terms in the density matrices of neutrinos and antineutrinos,  similar to what was discussed 
in~\cite{Dasgupta:2010ae} for three active flavors. 
The role  played by the neutrino self-interactions becomes  particularly evident at late postbounce times $t_0$,
because the matter background is lower and, 
therefore, the effective mixing angle $\theta_{13}$~\citep{EstebanPretel:2008ni,Duan:2008za} is larger than in the early cooling phase. 

 %--------------------------------------------
%--------------------------------------------

\newpage
%%%%%%%%%%%%%%%%%%%%%%%%%%%%%%%%%%%%%%%%%%%%%%%%%%%%%%%%%%%%%
%% Bibliography
%%%%%%%%%%%%%%%%%%%%%%%%%%%%%%%%%%%%%%%%%%%%%%%%%%%%%%%%%%%%%
\bibliographystyle{apj}
\bibliography{alpha_effect}{}

\end{document}